\documentclass[aps,pra,showpacs,twocolumn]{revtex4-1}%preprintnumbers,

\usepackage{graphicx}
\usepackage[dvipdfm,colorlinks]{hyperref}% [colorlinks=false][CJKbookmarks=true][pdftex]
\usepackage{amssymb}
\usepackage{amsmath}
\usepackage{amsfonts}

\begin{document}

\title{Effect of atomic distribution on cooperative spontaneous emission}
\author{Wei Feng}
\affiliation{Beijing Computational Science Research Center, Beijing 100084, China}
\author{Yong Li}
\affiliation{Beijing Computational Science Research Center, Beijing 100084, China}
\author{Shi-Yao Zhu}
\affiliation{Beijing Computational Science Research Center, Beijing 100084, China}
\date{\today }

\begin{abstract}
We study cooperative single-photon spontaneous emission from $N$ multilevel atoms for different atomic distributions in optical vector theory. Instead of the average approximation for interatomic distance or the continuum approximation (sums over atoms replaced by integrals) for atomic distribution, the positions of every atom are taken into account by numerical calculation. It is shown that the regularity of atomic distribution has considerable influence on cooperative spontaneous emission. For a small atomic sample (compared with radiation wavelength), to obtain strong superradiance not only needs the uniform excitation (the Dicke state) but also requires the uniform atomic distribution. For a large sample, the uniform atomic distribution is beneficial to subradiance of the Dicke state, while the influence of atomic distribution on the timed Dicke state is weak and its time evolution obeys exponential decay approximately. In addition, we also investigate the corresponding emission spectrum and verify the directed emission for the timed Dicke state for a large atomic sample.

\end{abstract}

\maketitle

\section{Introduction}

Superradiance is a well-known cooperative phenomenon that was first predicted in theory by Dicke in 1954~\cite{1}. An intriguing conclusion is that when $N$ identical atoms in the limit of a small dimension are uniformly excited by a single photon, the decay rate of the system is $N$ times that of the isolated atom. The cooperative effect is due to the exchange of real and virtual photons between atoms by the interaction field, and it is obvious that the distances between atoms play a key role in the cooperative spontaneous emission. The simplest theoretical model is the two-atom system~\cite{2,3,4,5}, where the dependence of collective decay rate and Lamb shift on the distance is clearly demonstrated. For the $N$-atom system, several models, such as atomic clouds with spherical, cubic and slab-shaped geometry, or atoms arranged as a straight line, have been studied with different approaches~\cite{6,7,8,9,10,11,12}. The manipulations of cooperative spontaneous emission can realize many interesting behaviors and potential applications~\cite{a1,a2,a3,a4}. Summary review on the cooperative spontaneous emission under certain approximations can be found in Refs.~\cite{13,14,15}.

Recently, cooperative spontaneous emission received renewed interest when the emission from an $N$-atom system excited by a single photon was considered~\cite{16,17,18,19,20,21,22,23,24,25,26,27,28,29,30}. With the development of experimental technology, it becomes possible to detect the single-photon superradiance~\cite{16}, which might yield new tools for storing quantum information and deepening our understanding on the physics of virtual processes~\cite{23}. The cooperative decay was observed more than 40 years ago~\cite{31}, but the direct experimental observation of the cooperative Lamb shift has been achieved just recently~\cite{32,33}. This breakthrough is stimulating the study on the effect of the virtual processes in the cooperative spontaneous emission. For a special initial state, i.e. the timed Dicke state, a directed emission with large atomic samples is predicted~\cite{16}. The dynamic problem involving the counter-rotating terms for a single excitation in a large atomic sample was studied in Ref.~\cite{20}, where, under several approximations, the author obtained the analytic conclusions indicating that the evolution of the timed Dicke state obeys the simple exponential decay.

In most of the previous papers on the cooperative spontaneous emission, the details of atomic positions are ignored and an average interatomic distance~\cite{11,12,a1,a2,a3,a4,13,14} or the continuum approximation~\cite{6,7,8,9,10,15,16,17,18,19,20,21,22,23,24,25,26,27} (sums over atoms replaced by integrals, actually, which is equivalent to the distance average) is taken. Furthermore, in most of recent papers about single-excitation cooperative spontaneous emission, the scalar photon theory which ignores the polarization and vector character of the field has been applied~\cite{16,17,18,19,20,21,22,23,24,25,26}. In our previous works~\cite{28,29}, we studied the same problem in optical vector theory but did not reckon in the electrostatic dipole-dipole interaction which dominates when the distance between atoms is much smaller than the wavelength. In another paper~\cite{30}, we have investigated the influence of the atomic distribution on the cooperative spontaneous emission for a simple model of three atoms with the electrostatic dipole-dipole interaction in optical vector theory. We found that the atomic distribution significantly influences the cooperative effects of the system, which leads us to suspect the universal applicability of the continuum approximation for multi-atom systems.

In this paper, we extend the study about the influence of the atomic distribution on the cooperative spontaneous emission from the three-atom system~\cite{30} to the $N$-atom one. The purpose of this paper is to illustrate that the spontaneous emission in random atomic distribution is very different from that in uniform distribution and the conclusions under the continuum approximation may be questionable for real experiments. The strong superradiance which approaches to the Dicke limit is hard to obtain in real experimental conditions where atoms are always distributed randomly. On the other hand, with the development of semiconductor quantum dots, the highly ordered arrays can be achieved in this artificial system~\cite{b1,b2} and its superradiance has been observed in experiment~\cite{34}. In the controllable artificial system, the cooperative effects can be studied in a regime which was difficult to achieve with real atoms~\cite{35,36,37}, and our results may be valuable to its experimental research.

\section{Model and Hamiltonian}
We consider a system consisting of $N$ identical multilevel atoms located at positions ${\mathbf{r}_j}$, $j = 1,...,N.$ The total Hamiltonian of the atoms and electromagnetic (EM) fields can be written as ($\hbar {\rm{ = }}1$)~\cite{13,30}

\begin{equation}
H = {H_0} + {H_{{\mathop{\rm int}} }} + {H_{d - d}},
\end{equation}
where
\begin{equation}
{H_0} = \sum\limits_{j = 1}^N {\sum\limits_l {{\omega _l}{{\left| l \right\rangle }_j}{{\left\langle l \right|}_j}} }  + \sum\limits_\mathbf{k} {{\omega _k}b_\mathbf{k}^\dag {b_\mathbf{k}}},\label{eq2}
\end{equation}
\begin{equation}
{H_{{\mathop{\rm int}} }} = \sum\limits_{j = 1}^N {\sum\limits_{l \ne m;\mathbf{k}} {{g_{\mathbf{k},lm}}{{\left| l \right\rangle }_j}{{\left\langle m \right|}_j}\left( {b_\mathbf{k}^\dag {e^{ - i\mathbf{k} \cdot {\mathbf{r}_j}}} + {b_\mathbf{k}}{e^{i\mathbf{k} \cdot {\mathbf{r}_j}}}} \right)} },\label{eq3}
\end{equation}
\begin{equation}
{H_{d - d}} = {\frac{1}{{4\pi {\varepsilon _0}}}}\sum\limits_{i < j} {\frac{{\mathbf{d}_i} \cdot {\mathbf{d}_j} - 3({\mathbf{d}_i} \cdot {{\mathbf{\hat r}}_{ij}})({\mathbf{d}_j} \cdot {{\mathbf{\hat r}}_{ij}})}{r_{ij}^3}}.\label{eq4}
\end{equation}
${H_0}$ is the unperturbed Hamiltonian of the atoms and fields, ${H_{{\mathop{\rm int}} }}$ is the interaction Hamiltonian between the atoms and the transverse fields, and ${H_{d - d}}$ is the electrostatic dipole-dipole interaction (also called instantaneous Coulomb interaction) between the atoms~\cite{13,15}. Here ${\omega _l}$ is the energy of the level $\left| l \right\rangle $, $b_\mathbf{k}^\dag $ (${b_\mathbf{k}}$) is the creation (annihilation) operator of the $\mathbf{k}$th EM mode with frequency ${\omega _k}$, and ${g_{\mathbf{k},lm}} = {\omega _{lm}}{d_{lm}}{(2{\varepsilon _0}{\omega _k}V)^{ - 1/2}}{\hat e_{\bf{\mathbf{k}}}} \cdot {{\bf{\mathbf{\hat d}}}_{lm}}$ is the coupling strength between the $\mathbf{k}$th EM mode with unit polarization vector ${\hat e_{\bf{\mathbf{k}}}}$ and the atomic transition between levels $\left| l \right\rangle $ and $\left| m \right\rangle $ with transition dipole moment ${\mathbf{d}_{lm}}{\rm{ = e}}\left\langle l \right|\mathbf{r}\left| m \right\rangle  = {d_{lm}}{{\bf{\mathbf{\hat d}}}_{lm}}$, of which ${d_{lm}}$ (assumed to be real) and ${{\bf{\mathbf{\hat d}}}_{lm}}$ are the magnitude and unit vector, respectively. The displacement between the $i$th and $j$th atoms is  ${\mathbf{r}_{ij}} \equiv {\mathbf{r}_j} - {\mathbf{r}_i} \equiv {r_{ij}}{{\bf{\mathbf{\hat r}}}_{ij}}$. In Eq.~(\ref{eq4}), ${\mathbf{d}_j} = \sum\nolimits_{lm} {{\mathbf{d}_{lm}}{{\left| l \right\rangle }_j}{{\left\langle m \right|}_j}} $ is the dipole moment operator of the $j$th atom. Here we have assumed that all the atoms are identical and similarly oriented~\cite{13}.

In order to take into account the counter-rotating terms and simplify the calculation, we introduce a unitary transformation $U{\rm{ = }}\exp (iS)$ ~\cite{38} with
\begin{equation}
S = \sum\limits_{j = 1}^N {\sum\limits_{l \ne m;\mathbf{k}} {{{{g_{\mathbf{k},lm}}{\xi _{k,lm}}} \over {i{\omega _k}}}} } {\left| l \right\rangle _j}{\left\langle m \right|_j}\left( {b_\mathbf{k}^\dag {e^{ - i\mathbf{k} \cdot {\mathbf{r}_j}}} - {b_\mathbf{k}}{e^{i\mathbf{k} \cdot {\mathbf{r}_j}}}} \right),
\end{equation}
where ${\xi _{k,lm}}{\rm{ = }}{{{\omega _k}} \mathord{\left/
{\vphantom {{{\omega _k}} {\left( {{\omega _k} + \left| {{\omega _{lm}}} \right|} \right)}}} \right.
\kern-\nulldelimiterspace} {\left( {{\omega _k} + \left| {{\omega _{lm}}} \right|} \right)}}$ and  ${\omega _{lm}} \equiv {\omega _l} - {\omega _m}$. In addition, we subtract the divergent free-electron self-energy ${E_{self}} =  - \sum\nolimits_{j,l \ne m} {\sum\nolimits_\mathbf{k} {\left( {{{{{\left| {{g_{\mathbf{k},lm}}} \right|}^2}} \mathord{\left/
{\vphantom {{{{\left| {{g_{k,\;lm}}} \right|}^2}} {{\omega _k}}}} \right.
\kern-\nulldelimiterspace} {{\omega _k}}}} \right){{\left| l \right\rangle }_j}{{\left\langle l \right|}_j}} }$  from the Hamiltonian. The effective Hamiltonian after the transformation can be written as~\cite{28,30}
\begin{align}
H^S  &  = {e^{iS}}H{e^{ - iS}} - {E_{self}}\nonumber\\
&  = {H'_0} + {H'_{{\mathop{\rm int}} }} + {H_{iv}} + {H_{d - d}} + O(g_{\mathbf{k},lm}^2),\label{eq6}
\end{align}
where
\begin{align}
{H'_0} &  = {H_0} + \sum\limits_{j = 1}^N \sum\limits_{l \ne m;\mathbf{k}} {{{{\left| {{g_{\mathbf{k},lm}}} \right|}^2}} \over {{\omega _k}}}\nonumber\\
& \times\left(\xi_{k,lm}^2 - {{{\omega_{lm}}} \over {{\omega_k}}}\xi_{k,lm}^2 - 2{\xi_{k,lm}} + 1\right){{\left| l \right\rangle }_j}{{\left\langle l \right|}_j},\label{eqA2}
\end{align}
\begin{align}
{H'_{{\mathop{\rm int}}}}=\sum\limits_{j = 1}^N{\sum\limits_{l < m;\mathbf{k}}{{g'_{\mathbf{k},lm}}({{{\left| l \right\rangle }_j}{{\left\langle m \right|}_j}b_\mathbf{k}^\dag {e^{-i\mathbf{k}\cdot{\mathbf{r}_j}}}+{{\left|m\right\rangle}_j}{{\left\langle l \right|}_j}{b_\mathbf{k}}{e^{i\mathbf{k}\cdot {\mathbf{r}_j}}}})}},
\end{align}
\begin{align}
{H_{iv}} & =  - \sum\limits_{i < j;\mathbf{k}} {\sum\limits_{l,l',m,m'} {{{2{g_{\mathbf{k},lm}}{g_{\mathbf{k},l'm'}}{\xi _{k,lm}}} \over {{\omega _k}}}(2 - {\xi_{k,l'm'}}){{\left| l \right\rangle }_i}{{\left\langle m \right|}_i}} }\nonumber\\
& \otimes {\left| {l'} \right\rangle _j}{\left\langle {m'} \right|_j}{e^{i\mathbf{k} \cdot {\mathbf{r}_{ij}}}}.
\end{align}
${H'_0}$ contains the non-dynamic Lamb shift for single atoms, i.e., the second term in Eq. (\ref{eqA2}), which is due to the counter-rotating terms. ${H'_{{\mathop{\rm int}}}}$ is the transformed interaction Hamiltonian describing the light-atom couplings, contains only the rotating wave terms (i.e., the terms associated with $\left| l \right\rangle \left\langle m \right|b_\mathbf{k}^\dag$ and $\left| m \right\rangle \left\langle l \right|{b_\mathbf{k}}$ where $\left| l \right\rangle$ is below $\left| m \right\rangle $), and ${g'_{\mathbf{k},lm}} = 2{g_{\mathbf{k},lm}}{{\left| {{\omega _{lm}}} \right|} \mathord{\left/
{\vphantom {{\left| {{\omega _{lm}}} \right|} {\left( {{\omega _k} + \left| {{\omega _{lm}}} \right|} \right)}}} \right.
\kern-\nulldelimiterspace} {\left( {{\omega _k} + \left| {{\omega _{lm}}} \right|} \right)}}$ is the transformed coupling strength. In contrast, the interaction Hamiltonian (\ref{eq3}) before the transformation describing the interaction between light and atoms contains both counter-rotating and rotating wave terms. The emerging term ${H_{iv}}$ describes the interatomic interaction due to the exchange of virtual photons (the counter-rotating terms). ${H_{d - d}}$ does not change its form after the unitary transformation because it commutes with $S$. $O(g_{\mathbf{k},lm}^2)$ contains terms of order $g_{\mathbf{k},lm}^3$ and higher, and it will be neglected.

\section{TIME EVOLUTION FOR SINGLE-ATOM-EXCITATION STATES}
Here we consider the weak excitation case in which only one of the atoms is in the first excited state and all others are in the ground state. Since the transformed interaction Hamiltonian ${H'_{{\mathop{\rm int}}}}$ only contains the rotating wave terms and the initial excitation is in the first excited state, the populations in higher atomic levels due to the counter-rotating terms can be neglected formally, and the multilevel atoms are reduced to effective two-level ($\left| e \right\rangle $ and $\left| g \right\rangle $) ones. In the interaction picture with respect to ${H'_0}$, the wave function at time t can be written as
\begin{equation}
\left| {\psi \left( t \right)} \right\rangle  = \sum\limits_{j = 1}^N {{\beta _j}(t)\left| {{e_j};0} \right\rangle }  + \sum\limits_{\bf{k}} {{\eta _\mathbf{k}}(t)\left| {G;{1_\mathbf{k}}} \right\rangle },\label{eq7}
\end{equation}
where $\left| {{e_j};0} \right\rangle \equiv \left|{{g_1}{g_2}\cdot \cdot  \cdot {e_j}\cdot  \cdot  \cdot {g_N}}\right\rangle\left|0\right\rangle$, and $\left| {G;{1_k}} \right\rangle  \equiv \left| {{g_1}{g_2} \cdot  \cdot  \cdot {g_N}} \right\rangle \left| {{1_{\bf{k}}}} \right\rangle $  with  $\left| 0 \right\rangle $ standing for the vacuum and $\left| {{1_{\bf{\mathbf{k}}}}} \right\rangle $  for one photon in the $\mathbf{k}$th  mode of the EM field. Substituting Eq. (\ref{eq7}) into the Schr\"{o}dinger equation yields the differential equations for ${\beta _j}(t)$ and ${\eta _\mathbf{k}}(t)$. Formally integrating the differential equation for ${\eta _\mathbf{k}}(t)$ with the initial value ${\eta_\mathbf{k}}(0)=0$, and then substituting into the differential equation for ${\beta _j}(t)$, under the Markov approximation, we find
\begin{equation}
{\dot \beta _i}(t) =  - \sum\limits_{j = 1}^N {{\Gamma _{ij}}{\beta _j}(t)},\label{eq8}
\end{equation}
where for $i = j$, ${\Gamma _{ii}} = {{{\gamma _0}} \mathord{\left/
 {\vphantom {{{\gamma _0}} 2}} \right.
 \kern-\nulldelimiterspace} 2} + i{\Delta _{eg}}$  with ${\gamma _0}{\rm{ = }}{{d_{eg}^2k_0^3} \mathord{\left/
 {\vphantom {{d_{eg}^2k_0^3} {\left( {3\pi {\varepsilon _0}} \right)}}} \right.
 \kern-\nulldelimiterspace} {\left( {3\pi {\varepsilon _0}} \right)}}$ the single-atom decay rate from  $\left| e \right\rangle $ to $\left| g \right\rangle $,  ${k_0} = {{{\omega _0}} \mathord{\left/
 {\vphantom {{{\omega _0}} c}} \right.
 \kern-\nulldelimiterspace} c}$ the wave number of resonant light, and ${\Delta _{eg}}$  the dynamic energy shift of single atoms. For $i \ne j$,
\begin{equation}
{\Gamma _{ij}} = {\frac{1}{2}}{\gamma _{ij}} + i{\Delta _{ij}},
\end{equation}
with
\begin{align}
\gamma _{ij}  &  = {\frac{3}{2}}{\gamma _0}\left\{ {{{\sin }^2}{\theta _{ij}}{\frac{\sin ({k_0}{r_{ij}})}{{k_0}{r_{ij}}}}} \right.\nonumber\\
& \left. { + \left( {1 - 3{{\cos }^2}{\theta _{ij}}} \right)\left[ {{{\cos ({k_0}{r_{ij}})} \over {{{({k_0}{r_{ij}})}^2}}} - {{\sin ({k_0}{r_{ij}})} \over {{{({k_0}{r_{ij}})}^3}}}} \right]} \right\},
\end{align}
\begin{align}
\Delta _{ij} &  = {3 \over 4}{\gamma _0}\left\{ { - {{\sin }^2}{\theta _{ij}}{{\cos ({k_0}{r_{ij}})} \over {{k_0}{r_{ij}}}}}\right.\nonumber\\
& \left. {+ (1 - 3{{\cos }^2}{\theta _{ij}})\left[ {{{\sin ({k_0}{r_{ij}})} \over {{{({k_0}{r_{ij}})}^2}}} + {{\cos ({k_0}{r_{ij}})} \over {{{({k_0}{r_{ij}})}^3}}}} \right]} \right\},\label{eq11}
\end{align}
where ${\theta _{ij}}$  is the angle between the dipole moment ${\mathbf{d}_{eg}}$ and the displacement ${\mathbf{r}_{ij}}$. The quantity ${\gamma _{ij}}$ accounts for the collective spontaneous emission effect and the quantity ${\Delta _{ij}}$ describes the vacuum-induced dipole-dipole interaction containing the electrostatic dipole-dipole interaction between atoms. The detailed calculations of above results can be found in Ref.~\cite{30}. Note that the two quantities ${\gamma _{ij}}$ and ${\Delta _{ij}}$ had emerged in the master equation in Refs.~\cite{11,12,13} where the Markov approximation had also been applied. For the Markov approximation to be valid, the atomic sample must not be too large such that the propagation effects, e.g. retardation, are negligible~\cite{24}.

The coefficients ${\Gamma_{ij}}$ in the differential equation (\ref{eq8}) constitute an $N \times N$  matrix. The problem reduces to the determination of complex eigenvalues ${\Gamma_n}$  and eigenstates $\left\vert \nu^{(n)}\right\rangle$ of the matrix ${\bf{\mathbf{\Gamma}}}$~\cite{17,22,28,29,30}. The eigenstates decay exponentially in the long-time limit. The decay rates of the eigenstates are given by $2{\mathop{\rm Re}\nolimits}({\Gamma_n})$, while ${\mathop{\rm Im}\nolimits}({\Gamma_n})$ are the corresponding Lamb shifts. For any initial state
\begin{equation}
\left| {\psi (0)} \right\rangle  = \sum\limits_{n = 1}^N {{c_n}} \left\vert \nu^{(n)}\right\rangle,\label{eq12}
\end{equation}
its time evolution is
\begin{equation}
\left| {\psi (t)} \right\rangle  = \sum\limits_{n = 1}^N {{c_n}{e^{ - {\Gamma _n}t}}\left\vert \nu^{(n)}\right\rangle}.\label{e13}
\end{equation}

\section{EFFECT OF ATOMIC DISTRIBUTION}
Here, we select two kinds of initial states of the system, the symmetric Dicke state~\cite{1}
\begin{equation}
\left| D \right\rangle  = {1 \over {\sqrt N }}\sum\limits_j {\left| {{e_j};0} \right\rangle},
\end{equation}
and the timed Dicke state~\cite{16}
\begin{equation}
\left| T \right\rangle  = {1 \over {\sqrt N }}\sum\limits_j {{e^{i{{\bf{k}}_0} \cdot {{\bf{r}}_j}}}\left| {{e_j};0} \right\rangle},\label{eq15}
\end{equation}
where ${{\bf{k}}_0}$ is the wave vector of the incident photon which prepares the single-excitation state. Note that the different initial states will result in different cooperative spontaneous emission, which also depends on the atomic distribution and the sample size. Next, by diagonalizing the matrix ${\bf{\Gamma}}$, we numerically analyze the influence of atomic distribution on the cooperative spontaneous emission for small and large samples, respectively.

\subsection{Effect of atomic distribution on superradiant limit in small sample}
First, we focus on a small sample, where the strong superradiance appears in the original theory of Dicke. The dimension of the atomic sample is smaller than the resonant wavelength, so that the additional position-dependent phase in Eq. (\ref{eq15}) is negligible and the timed Dicke state $\left|T\right\rangle$ reduces to the single-excitation Dicke state $\left| D \right\rangle$. In the following analysis, we will take the Dicke state as the initial state.
\begin{figure*}
\includegraphics[width=5.0cm,clip]{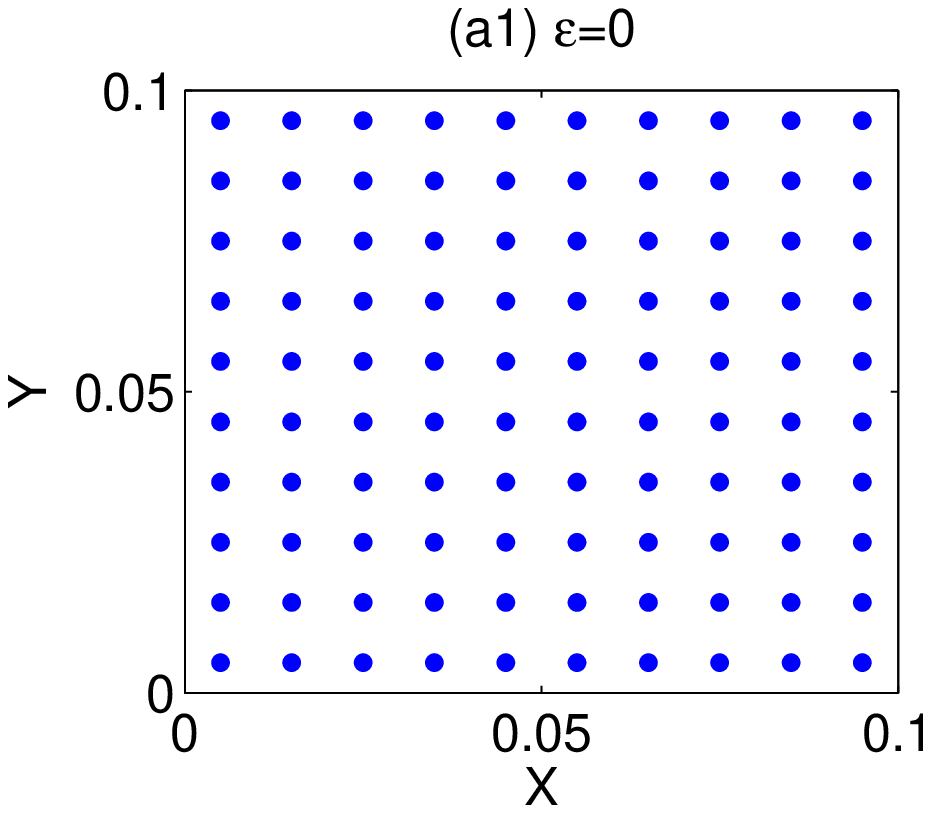}
\includegraphics[width=5.0cm,clip]{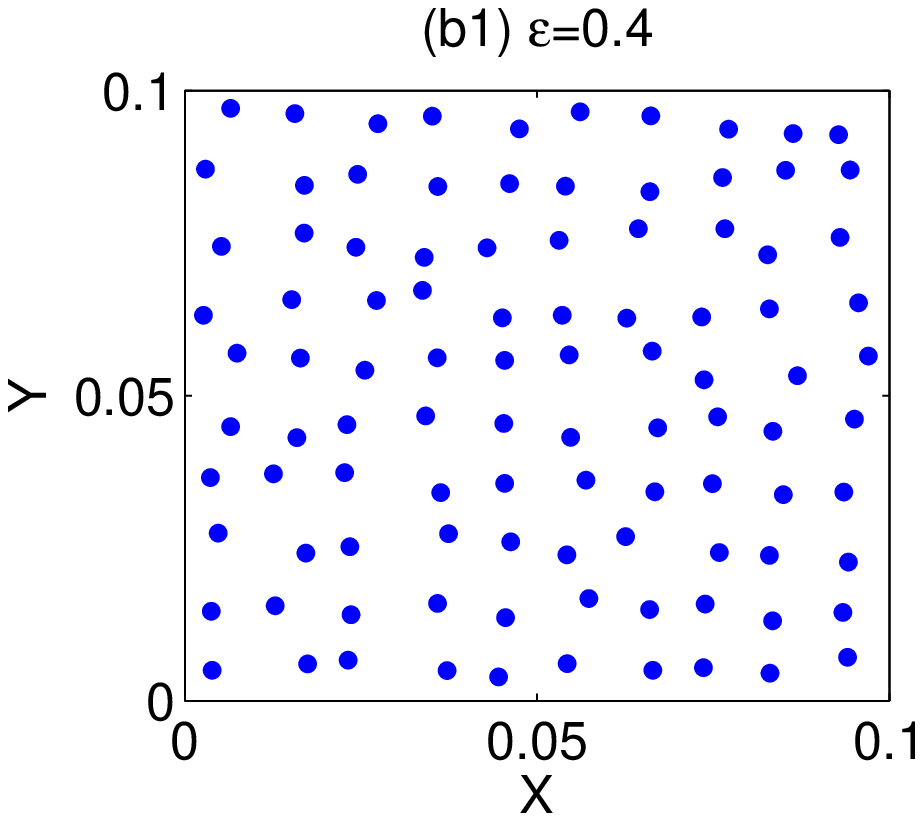}
\includegraphics[width=5.0cm,clip]{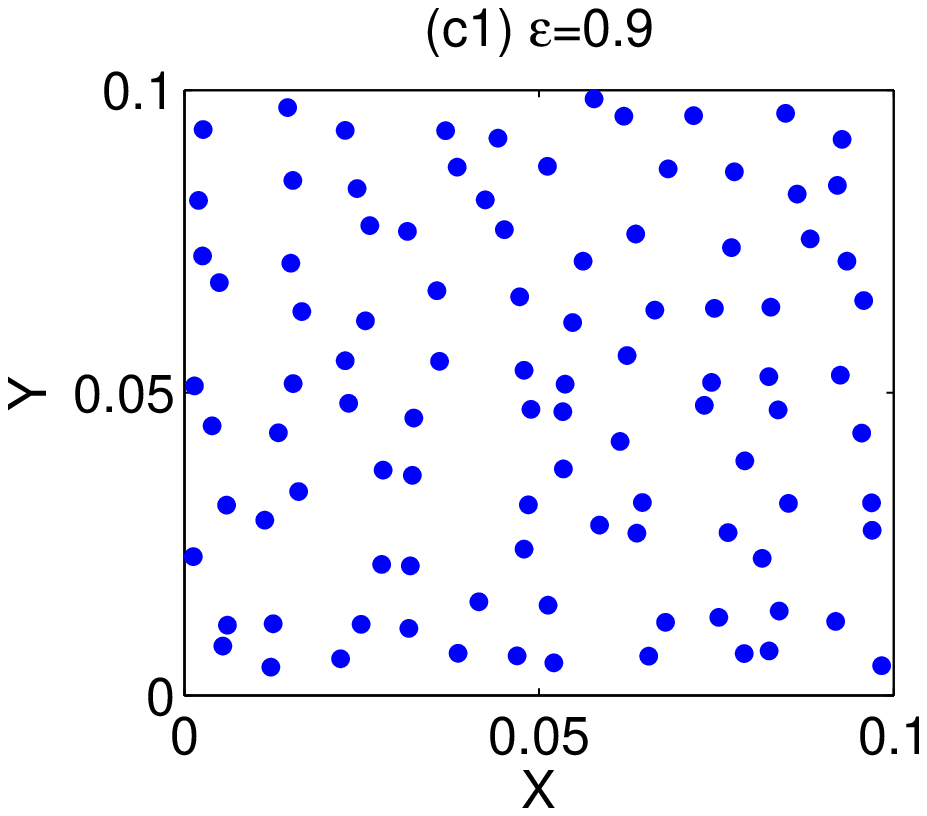}
%%%%%%%%
\includegraphics[width=5.0cm,clip]{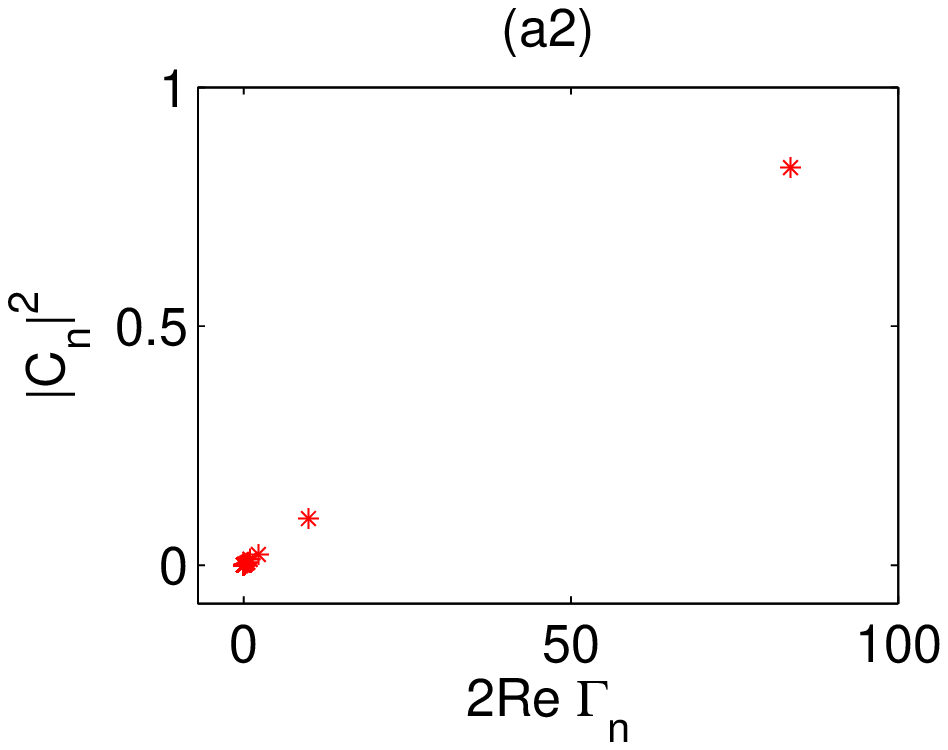}
\includegraphics[width=5.0cm,clip]{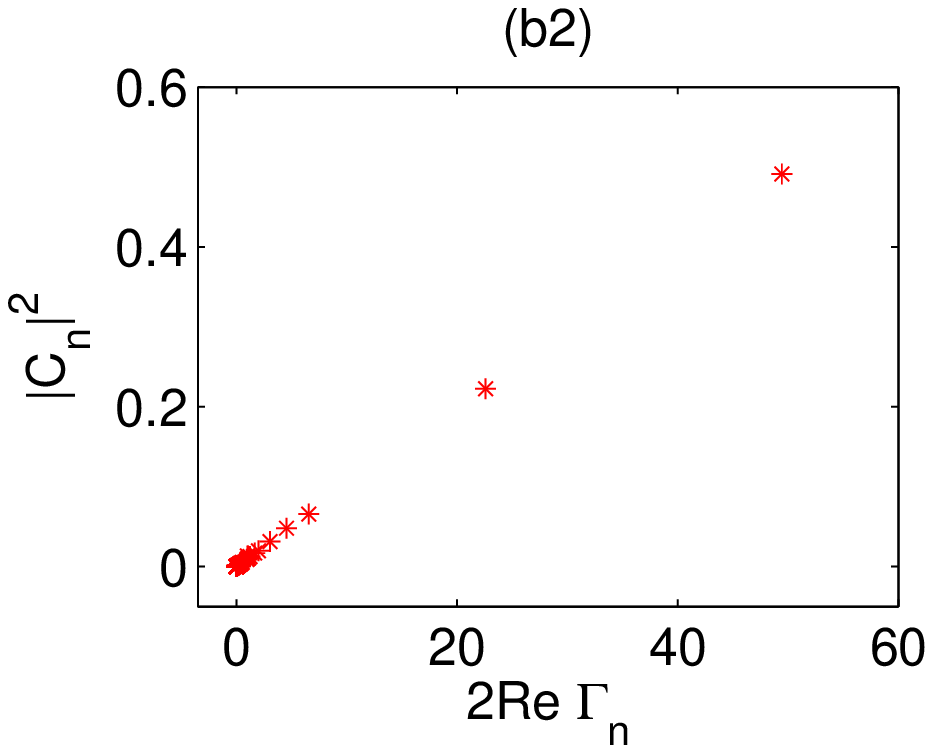}
\includegraphics[width=5.0cm,clip]{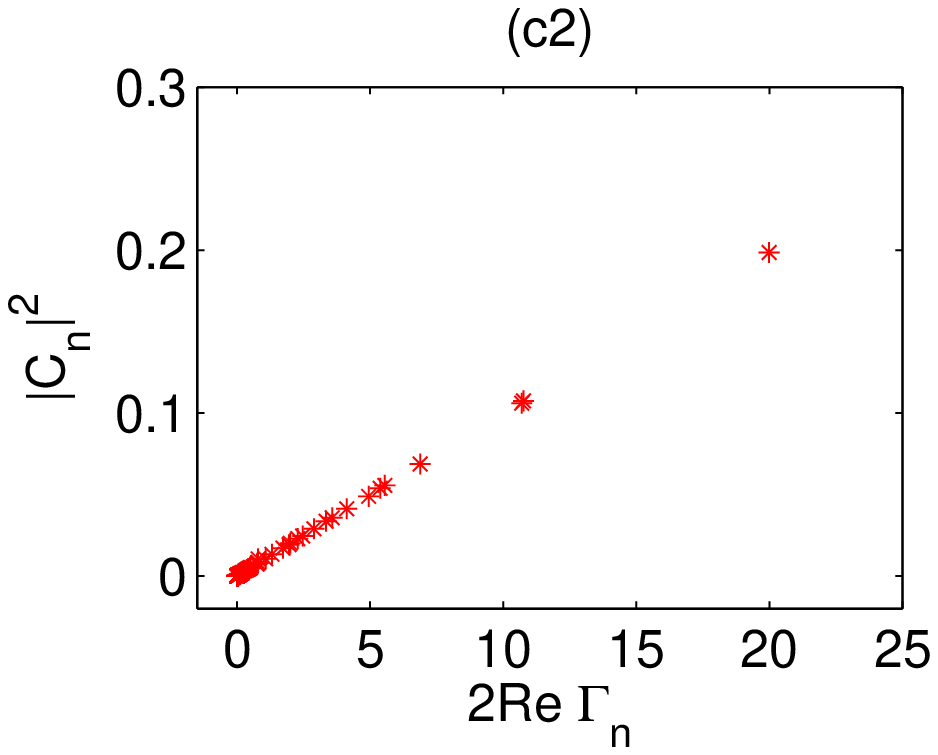}
%%%%%%
\caption{(Color online) Atomic distribution on a square (in units of ${\lambda_0}$) with different random parameter (a) $\varepsilon  = 0$, (b) $\varepsilon  = 0.4$, (c) $\varepsilon  = 0.9$ and their corresponding magnitude of the contribution of the different radiative eigenstates as functions of their decay rates  $2{\mathop{\rm Re}\nolimits} ({\Gamma _n})$ (in units of ${\gamma _0}$). Results are shown for the initial Dicke state $\left| D \right\rangle $  with the number of atoms $N = 100$.}
\label{fig1}
\end{figure*}
\begin{figure*}[htbp]
\includegraphics[width=5.0cm,clip]{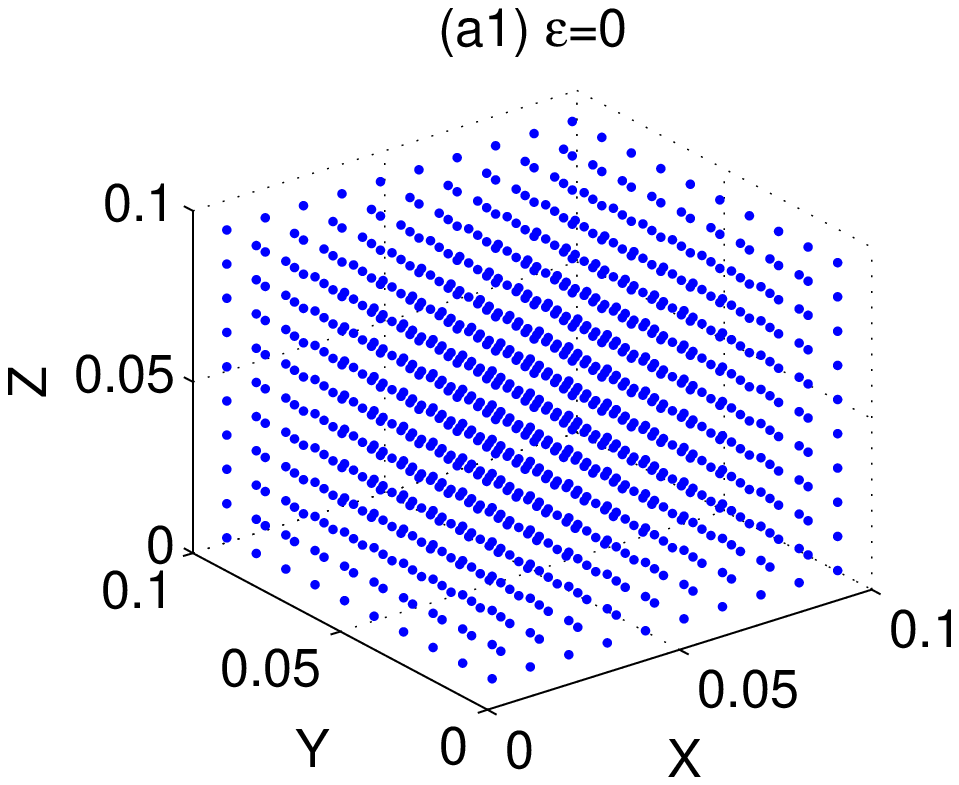}
\includegraphics[width=5.0cm,clip]{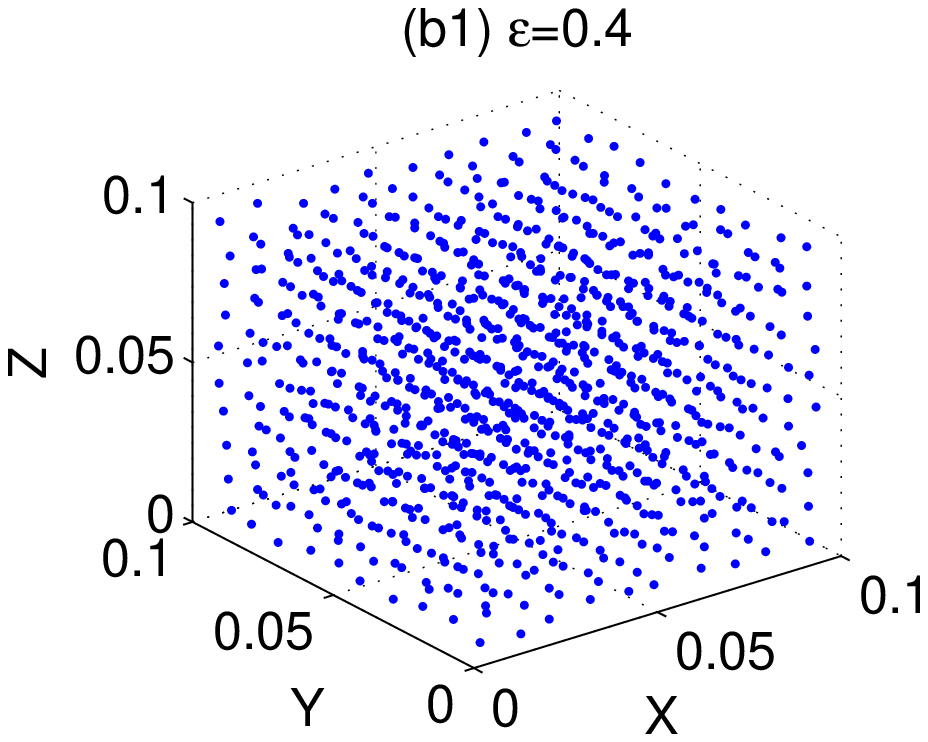}
\includegraphics[width=5.0cm,clip]{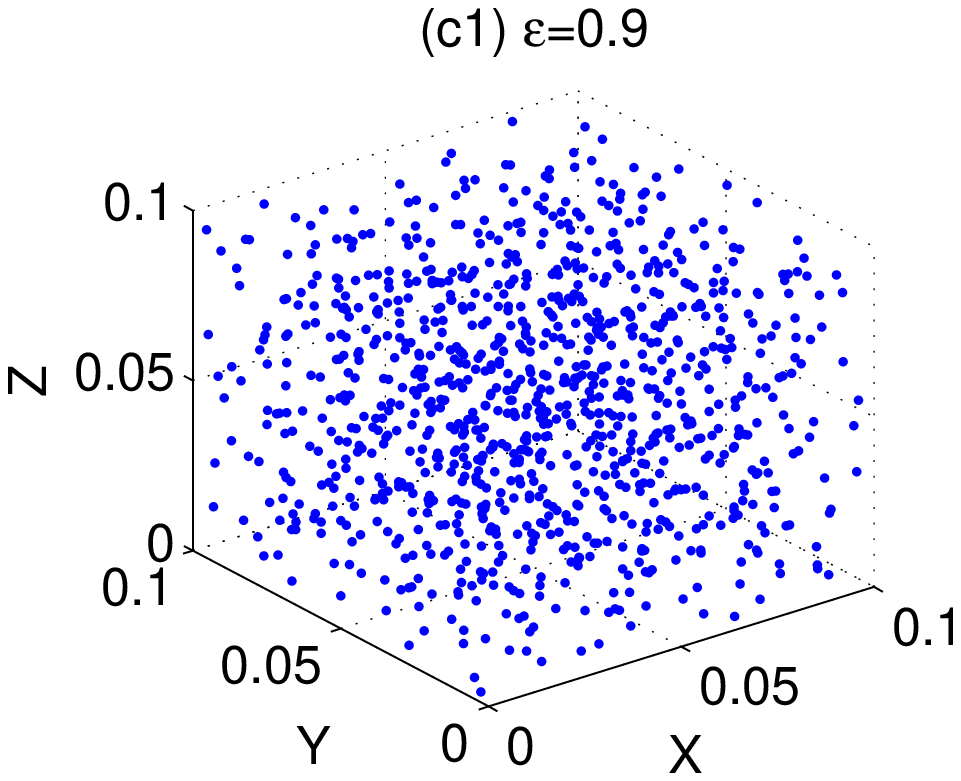}
%%%%%%%%
\includegraphics[width=5.0cm,clip]{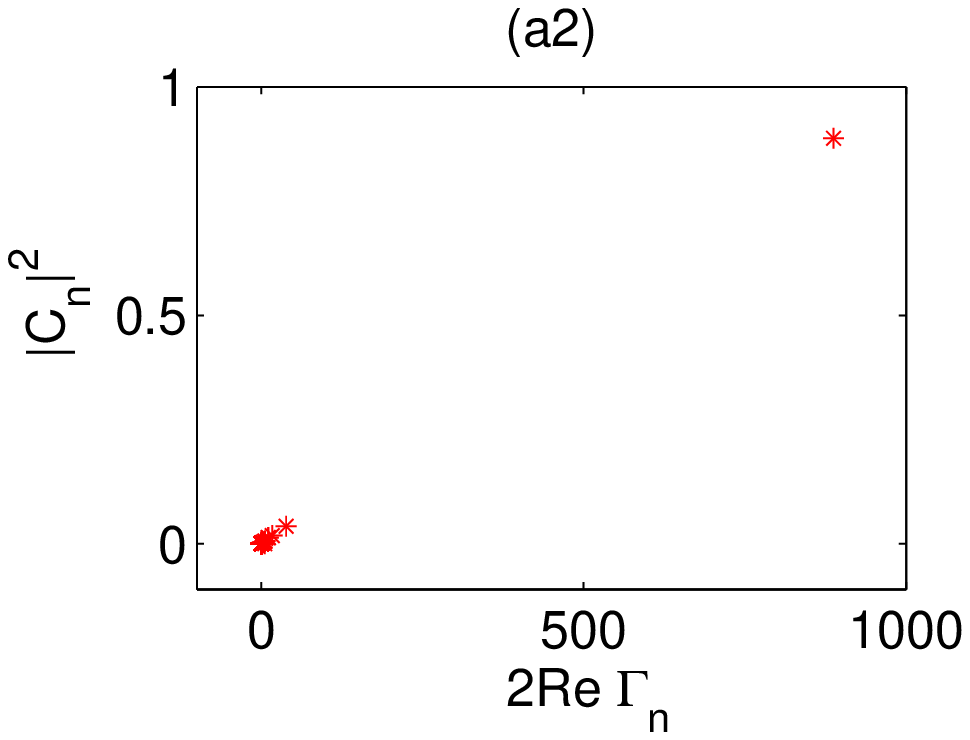}
\includegraphics[width=5.0cm,clip]{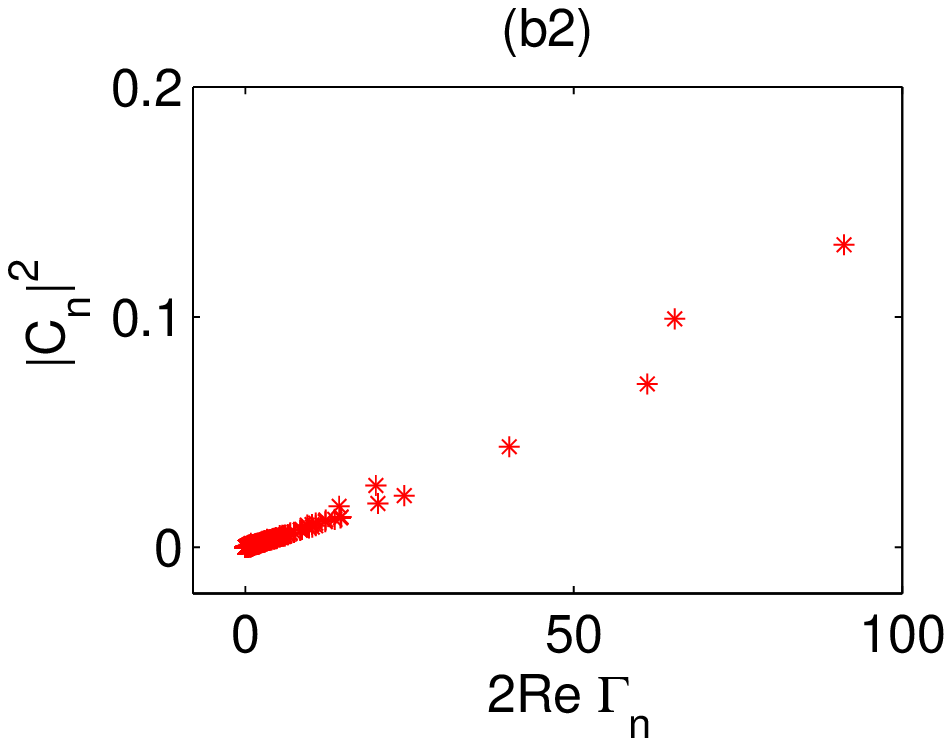}
\includegraphics[width=5.0cm,clip]{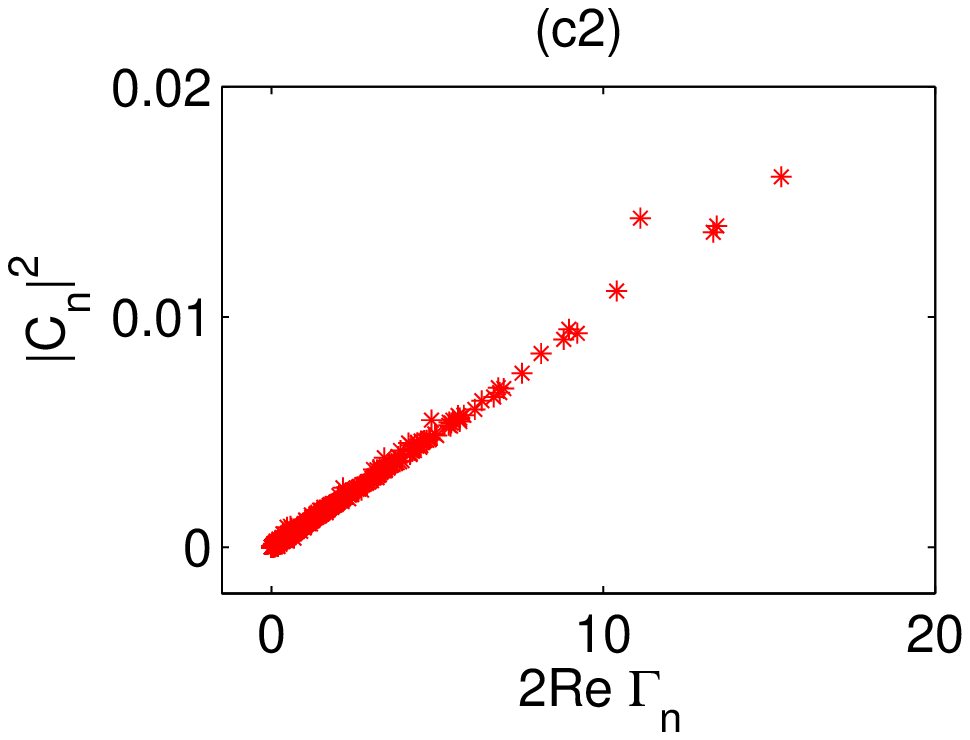}
%%%%%%
\caption{(Color online) Atomic distribution on a cube (in units of ${\lambda_0}$) with different random parameter (a) $\varepsilon  = 0$, (b) $\varepsilon  = 0.4$, (c) $\varepsilon  = 0.9$ and their corresponding magnitude of the contribution of the different radiative eigenstates as functions of their decay rates  $2{\mathop{\rm Re}\nolimits} ({\Gamma _n})$ (in units of ${\gamma _0}$). Results are shown for the initial Dicke state $\left| D \right\rangle $  with the number of atoms $N = 1000$.}
\label{fig2}
\end{figure*}

We select two different geometry cases, a square and a cube. For each case, we made the atomic distribution from totally random to slight disturbance and finally, completely regular. To produce the random or slightly disturbed positions of atoms, we first fix atoms regularly in their own positions [see Figs.~\ref{fig1}(a1) and~\ref{fig2}(a1)], where the lattice constant is denoted by $a$, then we add a random displacement $\varepsilon a[{\rm{rand(}}{{\rm{r}}_j}{\rm{) - 0}}{\rm{.5}}]$  on each atom [see Figs.~\ref{fig1}(b1),~\ref{fig1}(c1),~\ref{fig2}(b1) and~\ref{fig2}(c1)], where $0 \le \varepsilon < 1$  and ${\rm{rand(}}{{\rm{r}}_j}{\rm{)}}$  is a random number between 0 and 1. So that we can adjust the randomness of atomic distribution by controlling the parameter $\varepsilon $. Note that the dipole-dipole interaction is divergent as ${r^{-3}}$  when the atomic distance approaches to zero [see Eq.(\ref{eq11})]. Actually, this is not physical and we can make our calculation valid by forbidding the small atomic distance which has the same order with the Bohr radius. In Figs.~\ref{fig1} and~\ref{fig2}, the side lengths of the square and the cube are both  $0.1{\lambda_0}$ with $N = 100$ and $N = 1000$, respectively, so the lattice constants are both $a = 0.01{\lambda _0}$. We estimate that the transition wavelength ${\lambda_0} = 500\operatorname{nm}$  and the Bohr radius ${r_0} = 0.053\operatorname{nm}$, so that if we control $\varepsilon  \le 0.9$, the nearest atomic separation after the random displacing $\varepsilon a[{\rm{rand(}}{{\rm{r}}_j}{\rm{) - 0}}{\rm{.5}}]$ will be larger than $10{r_0}$.

In the second row of Figs.~\ref{fig1} and ~\ref{fig2}, we plot the structures of the radiative eigenstates for the corresponding atomic distributions shown in the first row. The horizontal axis represents the decay rates ${\gamma _n}{\rm{ = }}2{\mathop{\rm Re}\nolimits} ({\Gamma _n})$ of the eigenstates $\left\vert \nu^{(n)}\right\rangle $ and the vertical axis is their weights ${\left| {{c_n}} \right|^2}$  with the initial Dicke state $\left| D \right\rangle  = \sum\limits_{n = 1}^N {{c_n}} \left\vert \nu^{(n)}\right\rangle $. We can see that in Figs.~\ref{fig1}(a2) and ~\ref{fig2}(a2), there is only one dot appearing with large ${\left| {{c_n}} \right|^2}$ and with its decay rate close to the superradiant limit $N{\gamma_0}$, while the dots belonging to all other modes have negligible contribution for the initial Dicke state. This means, the Dicke state is a approximate radiative eigenstate with the large decay rate close to superradiant limit $N{\gamma _0}$  for the small atomic sample under the regular distribution. However, if the atoms are distributed randomly, the Dicke state is composed of many different eigenstates and their decay rates are all much smaller than $N{\gamma_0}$ , see Figs.~\ref{fig1}(b2),~\ref{fig1}(c2),~\ref{fig2}(b2) and ~\ref{fig2}(c2).

To characterize the time evolution of the system more clearly, from Eq.(\ref{e13}) we calculate the time-dependent population in all atomic excited state
\begin{equation}
P\left( t \right) = \sum\limits_j {{{\left| {\left\langle {{e_j};0\left| {\psi (t)} \right\rangle } \right.}\right|}^2}}.
\end{equation}
\begin{figure}[b]
\includegraphics[width=7.0cm,clip]{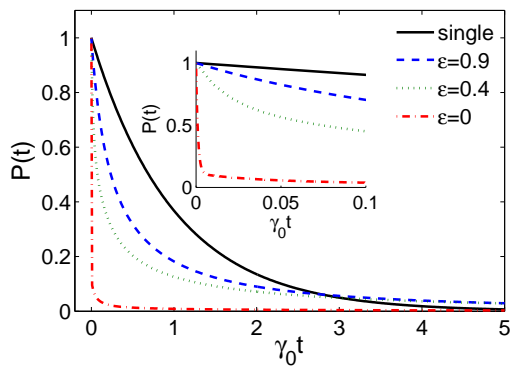}
\caption{(Color online) The upper-state population  $P\left( t \right)$ for the small atomic sample given in Fig.~\ref{fig2} with different atom-distribution random parameter $\varepsilon = 0$ (dash dot), $\varepsilon  = 0.4$ (dot) and $\varepsilon  = 0.9$ (dash) and the single-atom case (solid). The insert is the amplification of the little range. Here $N=1000$ and the initial state is the symmetric Dicke state $\left| D \right\rangle$.}
\label{fig3}
\end{figure}
The populations corresponding to different atomic distributions are plotted in Fig.~\ref{fig3}. For comparison, we also plot the single-atom decay (see the black solid line). From this figure, we can see the emission of small atomic sample for the initial Dicke state is superradiant whatever the distribution is. But, when the atoms are distributed regularly (see the red dashed dotted line), the decay of the system will be much faster than that in the case of random atomic distribution (see the blue dashed line and the green dotted line). This result coincides with the structural analysis of the radiative eigenstates shown in Fig.~\ref{fig2}.

Here we try to give a physical interpretation of the above numerical results. The cooperative effect of spontaneous emission is influenced by the dipole-dipole interaction between atoms which has the ${r^{-3}}$ behavior in the small atomic distance. If the atoms are distributed randomly, some part of them which are close to each other may converge together as a small subsystem. So the atomic ensemble is split up into multiple subsystems whose weak cooperative effects are restricted in their own interior. However, when the atoms are distributed uniformly, the photons can be uniformly exchanged among them and as a whole they can display the strong collective spontaneous emission. In Refs.~\cite{28,29}, the electrostatic dipole-dipole interaction is neglected unreasonably, then the interactions between atoms are less sensitive to the distances and the strong superradiance is still obtained under the random distribution. Note that the effect of regular atomic distribution is universal for different geometries (see Figs.~\ref{fig1} and ~\ref{fig2}), therefore, we only choose the cubic geometry in the below analysis.

\subsection{Effect of atomic distribution on cooperative spontaneous emission for large sample}
For large multi-atom samples, the timed Dicke state is superradiant, while the symmetric Dicke state is subradiant~\cite{22,23}. This previous conclusion is also based on the continuum approximation of atomic distribution. Here, we discuss the influence of atomic distribution (random or regular distribution) on the cooperative spontaneous emission of large atomic samples.

\begin{figure*}[htbp]
\includegraphics[width=5.0cm,clip]{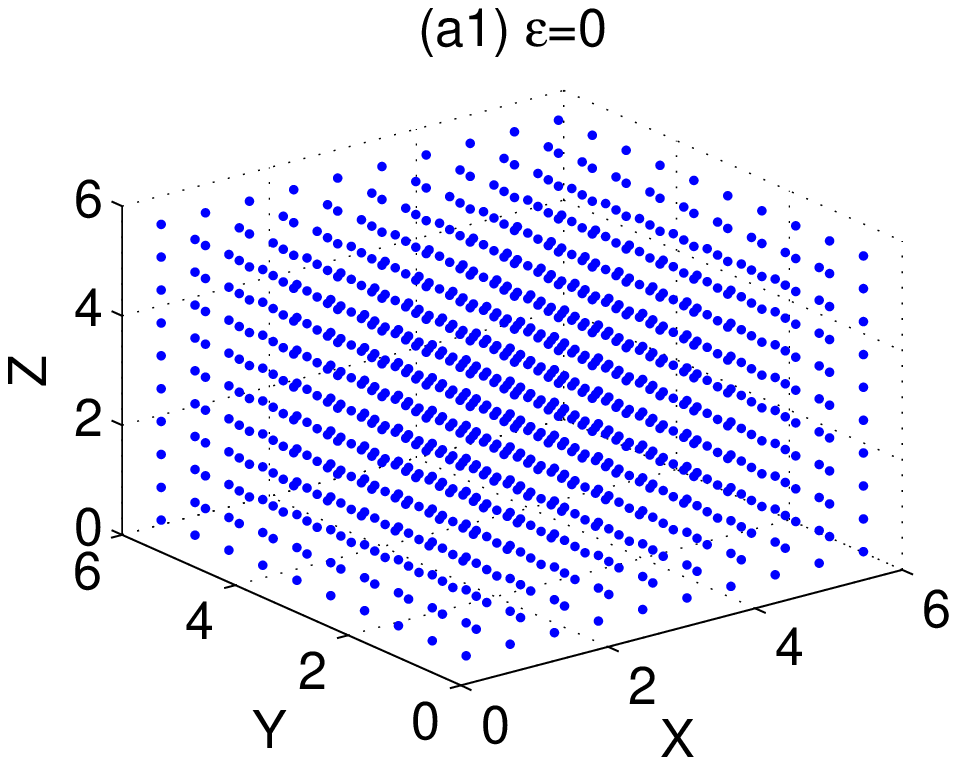}
\includegraphics[width=5.0cm,clip]{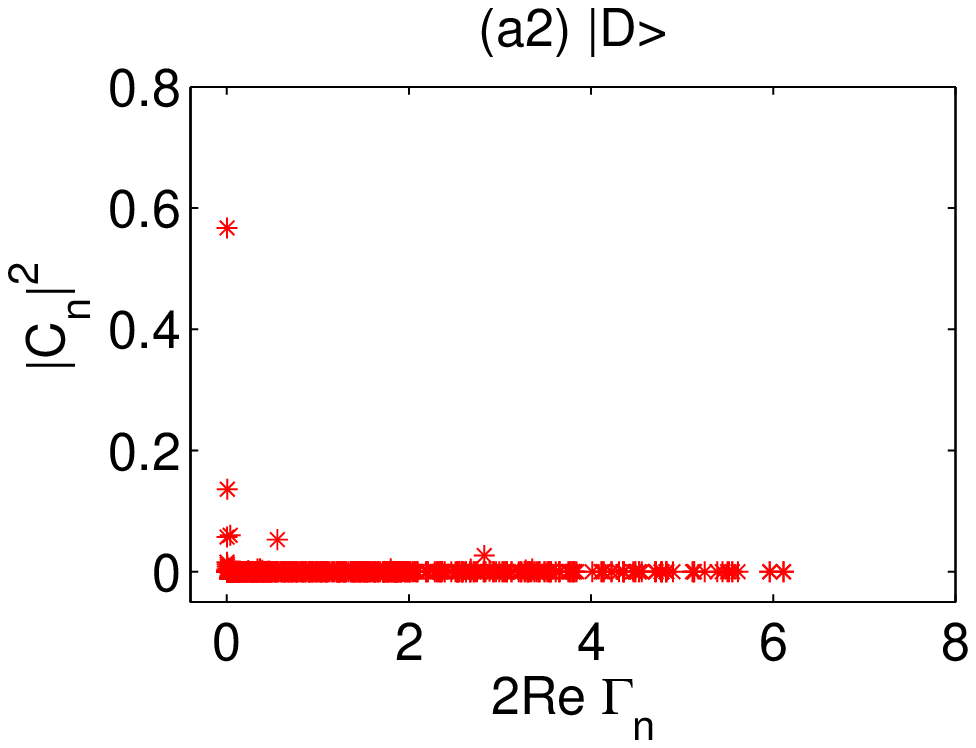}
\includegraphics[width=5.0cm,clip]{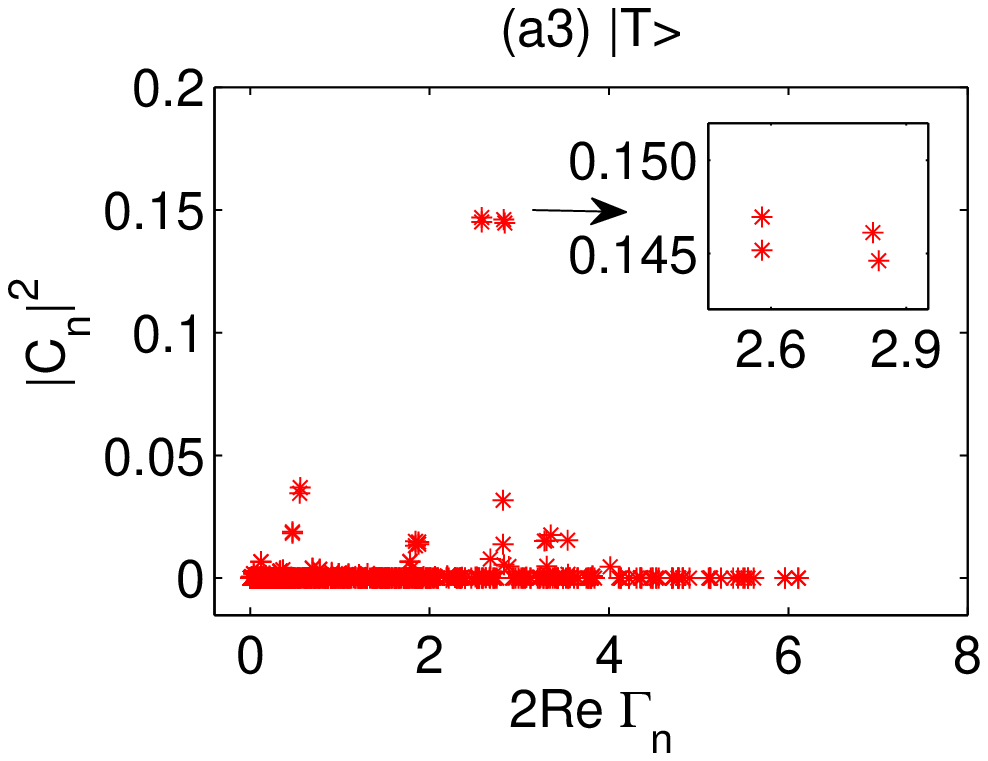}
%%%%%%%%
\includegraphics[width=5.0cm,clip]{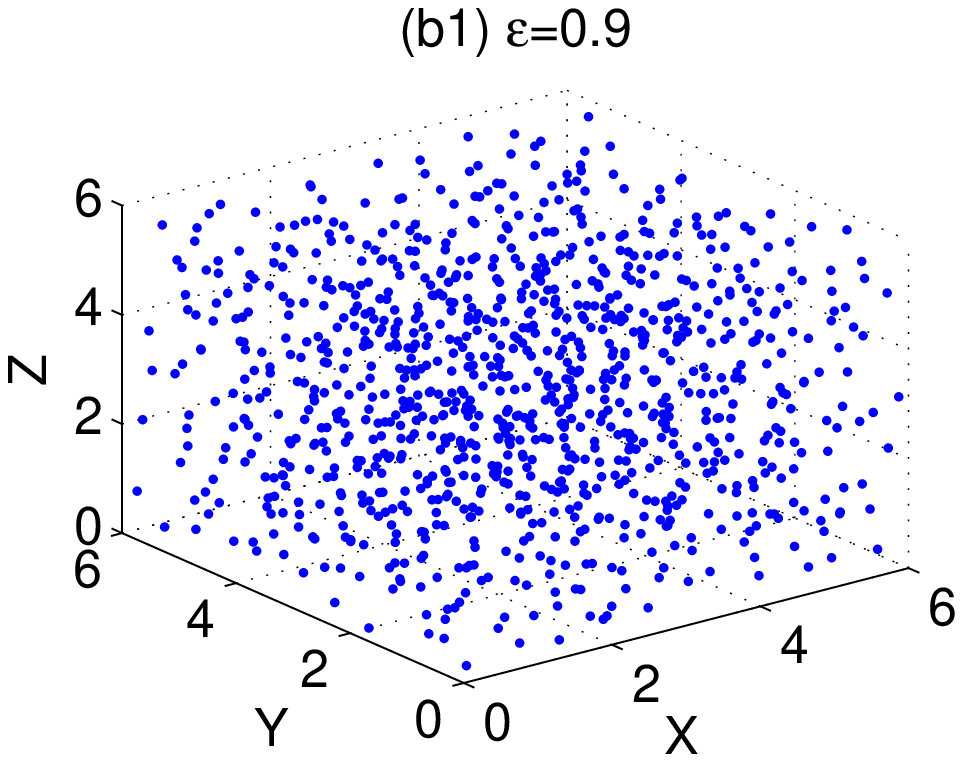}
\includegraphics[width=5.0cm,clip]{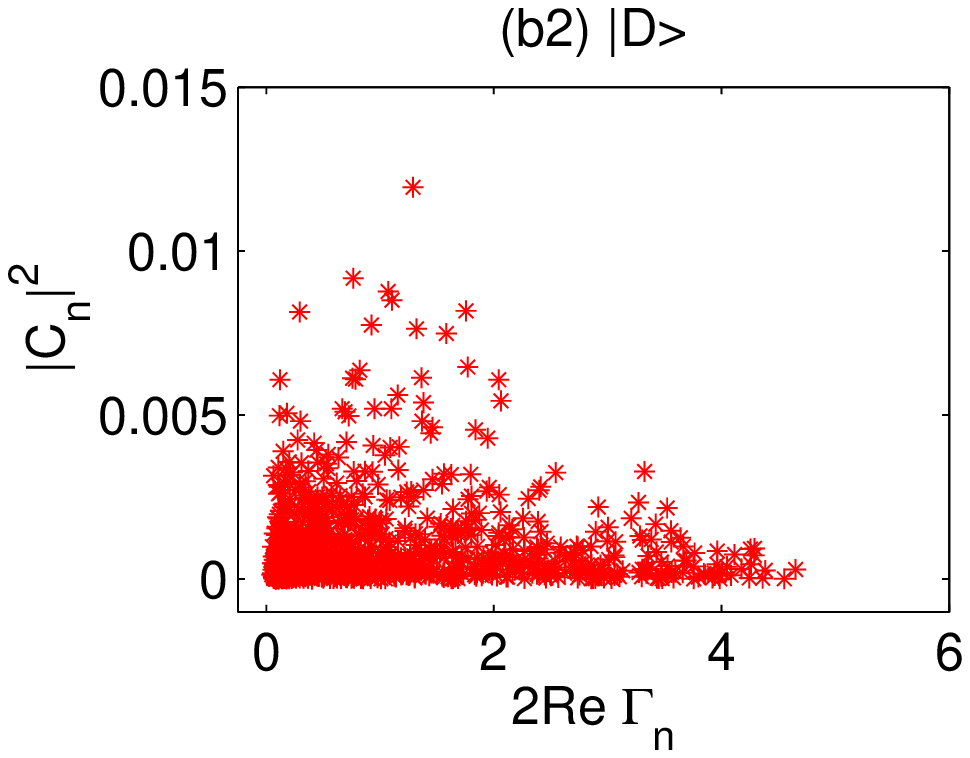}
\includegraphics[width=5.0cm,clip]{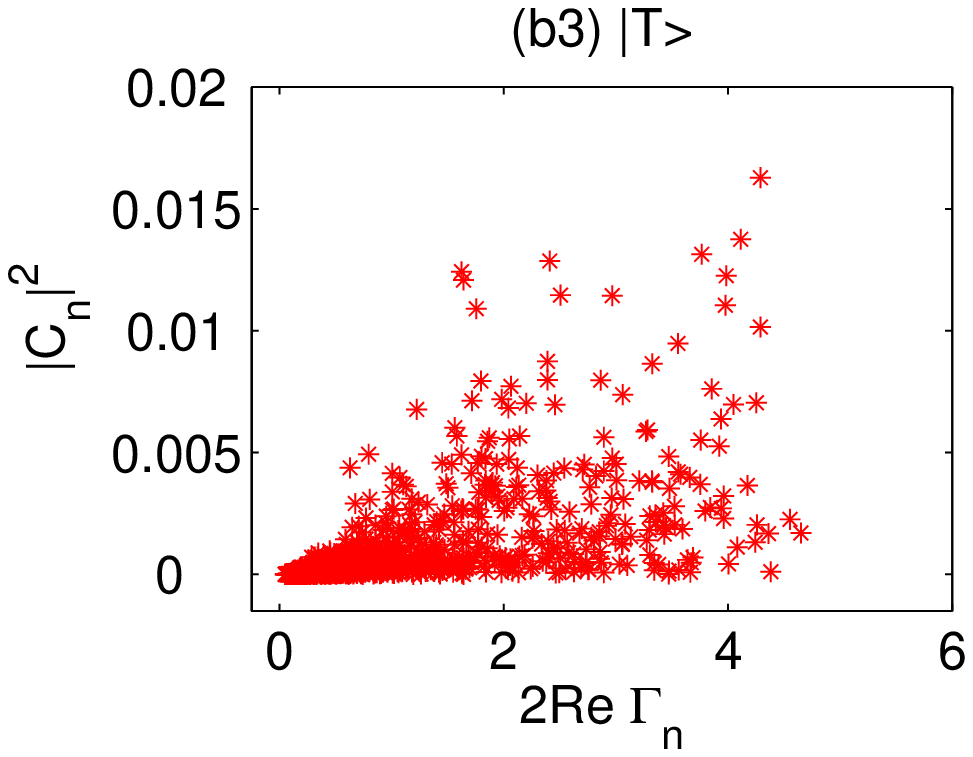}
%%%%%%
\caption{(Color online) Atomic distribution on a cube (in units of ${\lambda_0}$) with different random parameter (a1) $\varepsilon = 0$ and (b1) $\varepsilon = 0.9$. (a2) and (b2) [(a3) and (b3)] show the corresponding magnitude of the contribution of the different radiative eigenstates as functions of their decay rates  $2{\mathop{\rm Re}\nolimits} ({\Gamma_n})$ (in units of ${\gamma_0}$) for the initial Dicke state $\left| D \right\rangle $ (the timed Dicke state $\left| T \right\rangle $). In (a3), the insert is the amplification of the principal range. Here the number of atoms is $N = 1000$.}
\label{fig4}
\end{figure*}

In Fig.~\ref{fig4}, we plot the regular and random atomic distributions and their corresponding structures of the radiative eigenstates with the initial Dicke and timed Dicke states. In the cases of random large samples, both the Dicke and timed Dicke states consist of many components of different eigenstates, as shown in Figs.~\ref{fig4}(b2) and~\ref{fig4}(b3). Instead, for regular large samples, the Dicke state largely projects to a few eigenstates with one of them holding nearly $60{\rm{\% }}$ probability, and the timed Dicke state also mostly projects to a few eigenstates whose corresponding ${\left| {{c_n}} \right|^2}$ are biggish and similar, see Figs.~\ref{fig4}(a2) and~\ref{fig4}(a3), respectively.

\begin{figure*}[htbp]
\includegraphics[width=6.5cm,clip]{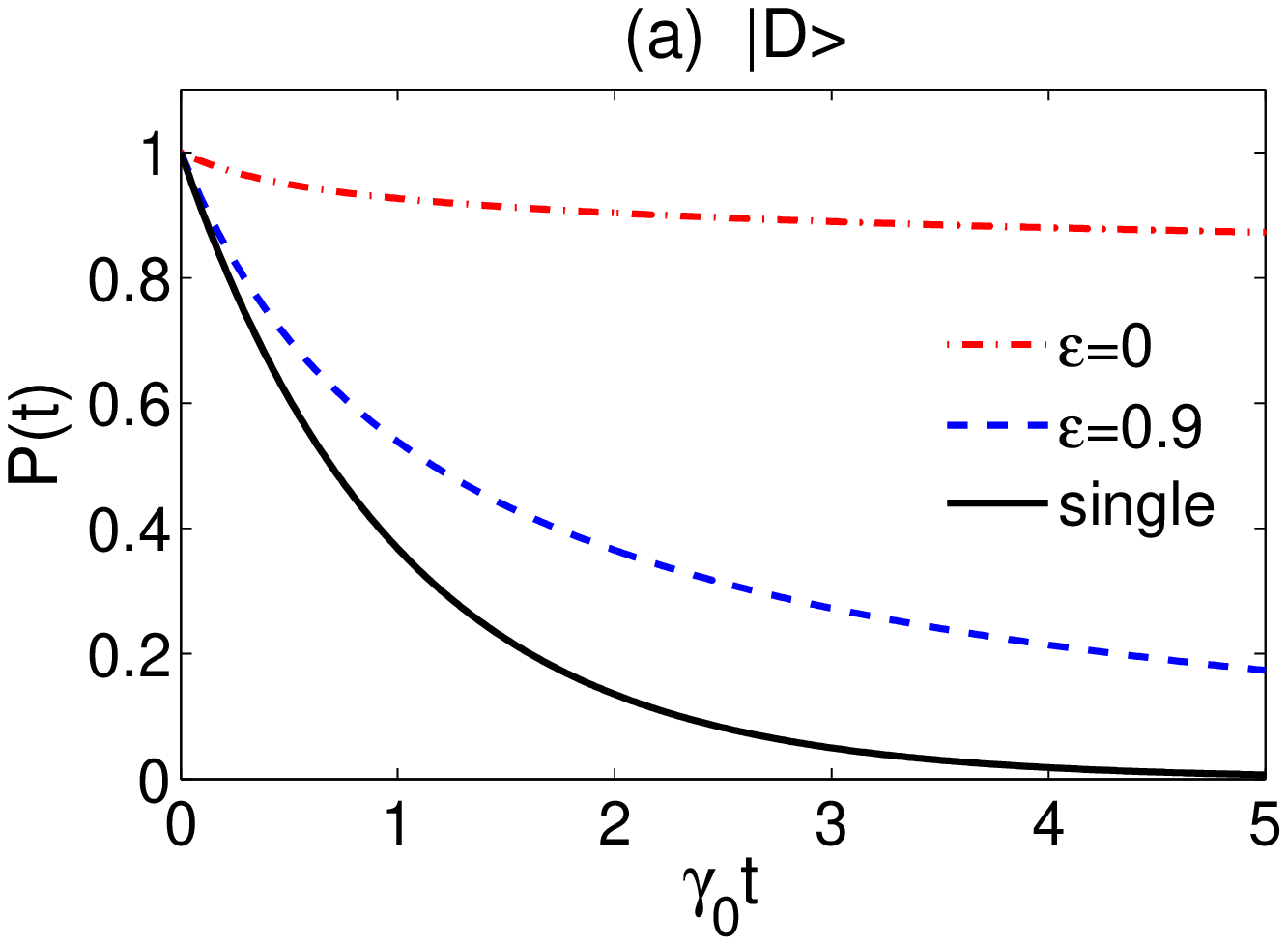}
\includegraphics[width=6.5cm,clip]{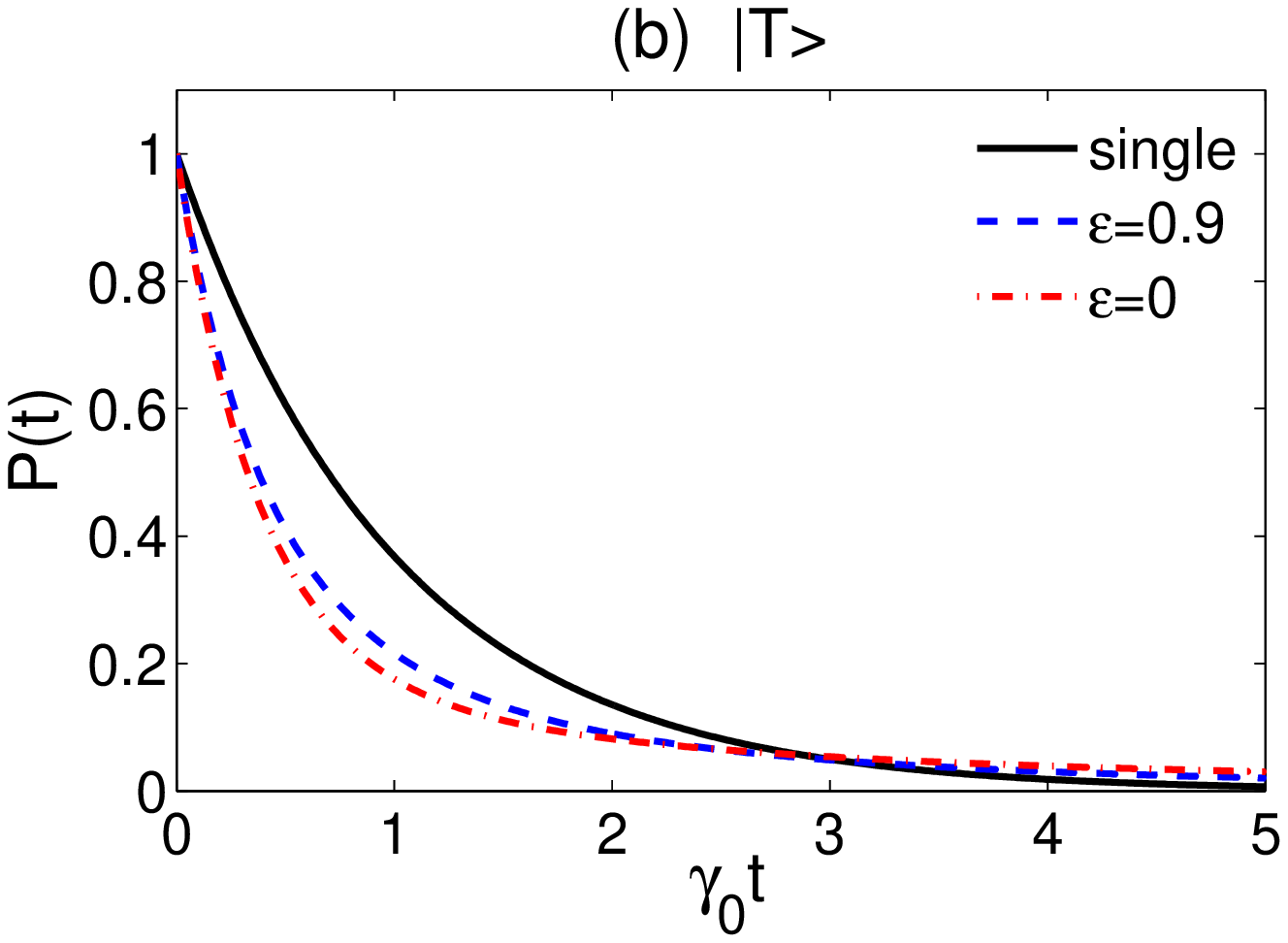}
\caption{(Color online) The upper-state population $P\left( t \right)$ for the large atomic sample given in Fig.~\ref{fig4} with different atom-distribution random parameter $\varepsilon  = 0$ (dash dot) and $\varepsilon  = 0.9$ (dash) and the single-atom case (solid). Here the initial states are (a) the symmetric Dicke state $\left| D \right\rangle $ and (b) the timed Dicke state $\left| T \right\rangle $.}
\label{fig5}
\end{figure*}

We also plot the probability that atoms are still in the excited states for the two cases of random (blue dashed lines) and regular (red dashed dotted lines) distributions for the Dicke and timed Dicke states in Fig.~\ref{fig5}, wherein the solid lines mean the decay of a single-atom for comparison.

From Fig.~\ref{fig5}(a) we can see that the Dicke state is surely a subradiant state and its decay is slower than single-atom one whatever the atomic distribution is. This result is similar to that in Refs.~\cite{22,23}, where the Dicke state is not a good trapped state and its excitation will be slowly emitted due to the virtual processes. However, if the atoms are distributed regularly, the decay of the excitation will be slower than that in the case of general random distribution obviously. Of course, it is not realistic to control the atoms in regular distribution for the sample of low-pressure gas in most experiments. However, the semi-conduct quantum dot is the burgeoning and hopeful system to control the distribution and compass the purpose of storing the photon by subradiance with the Dicke state in large sample. Here, our results demonstrate that the regular distribution is beneficial to realize the photon storage.

In Fig.~\ref{fig5}(b), we plot the upper-state population for the initial timed Dicke state. Here, we want to recall Fig.~\ref{fig4}(a3) first, where there are four eigenstates that dominate the evolution of the timed Dicke state. It clearly demonstrates that the timed Dicke state is not an approximate eigenstate for large atomic sample. However, due to the real parts of the four eigenvalues are approximately equal, the timed Dicke state can be approximately written in an exponentially decaying form even if it is not an eigenstate. This result is coincident with the conclusion in Refs.~\cite{20,21,22,23}. From Fig.~\ref{fig5}(b), we can see the approximately exponentially decaying behavior indeed. In addition, the difference of the evolved upper-state populations between the regular and random distributions for the timed Dicke state is small, and the decays under the two distributions are both superradiant.

\section{DIRECTED EMISSION}
In our previous works ~\cite{29,30}, we studied the total spectrum of cooperative spontaneous emission which is the average of the spectra detected in each direction. Here we investigate the directed properties of the spontaneous emission for the initial timed Dicke state~\cite{16} and the influence of atomic distribution on the emission spectrum. The spectrum detected by the detector at position  $\mathbf{R} \equiv R\mathbf{\hat R}$ is given by Refs.~\cite{29,30}
\begin{equation}
{S_\mathbf{R}}({\omega _k}) \propto {\left| {\mathbf{\hat R} \times ({{\mathbf{\hat d}}_{eg}} \times \mathbf{\hat R})\sum\limits_j {\sum\limits_n {{{{e^{i{\bf{k}} \cdot {{\bf{R}}_j}}}{c_n}\nu_j^{(n)}} \over {{\Gamma_n} - i({\omega_k} - {\omega_{0}})}}}}}\right|^2},
\end{equation}
where ${\mathbf{R}_j}\equiv\mathbf{R}-{\mathbf{r}_j}\equiv{R_j}{\mathbf{\hat R}_j}$, ${c_n}$ is the coefficient in Eq.~(\ref{eq12}), and $\nu_j^{(n)}=\left\langle {{e_j};0}\right|\left. {{\nu^{(n)}}} \right\rangle $. Note that the detector is far away from the atoms in experiment ($R \gg {\lambda _0}$). The optical mode whose wavevector is not parallel to $\mathbf{\hat R}$ leads to negligible contributions to the detector, so $\mathbf{\hat k}$ is replaced by $\mathbf{\hat R}$. Meanwhile, as the detuning ${\delta_k} \equiv {\omega_k} - {\omega_{0}}$  is far less than the resonant frequency ${\omega_{0}}$, we can replace $k$ by ${k_0}$ in the exponential term. Based on these approximations, we have ${e^{i{\bf{k}} \cdot {{\bf{R}}_j}}} = {e^{i{\bf{k}} \cdot \left( {{\bf{R}} - {{\bf{r}}_j}} \right)}} = {e^{i{k_0}R}}{e^{-i{k_0}{\bf{\hat R}} \cdot {{\bf{r}}_j}}}$.

\begin{figure}[t!]
\includegraphics[width=6.0cm,clip]{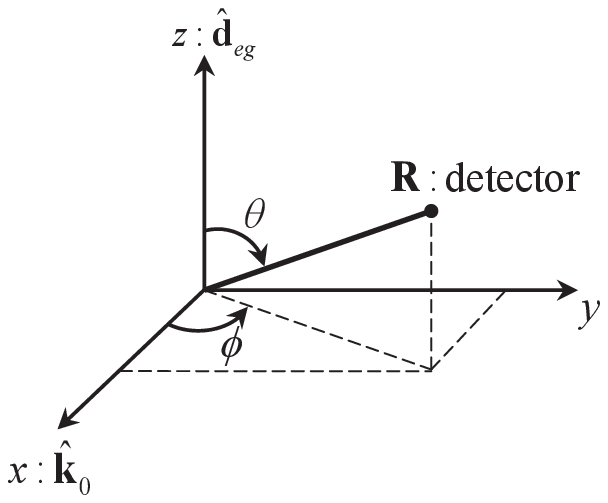}
\caption{The schematic description of the position of the detector.}%
\label{fig6}%
\end{figure}

Next, we select the direction of the atomic dipole moment ${{\bf{d}}_{eg}}$ in the $z$ direction, and assume the wave vector of the incident light which prepares the timed Dicke state along the $x$ direction. The position of the detector in the coordinate system is shown in Fig.~\ref{fig6}, where $\theta$ is the angle between the detector and the $z$ direction, $\phi$ is the azimuthal angle of the vector. In the selected coordinate system, the expression of spectrum becomes
\begin{equation}
{S_{\mathbf{\hat R}}}({\omega _k})\propto{\left|{\sin \theta \sum\limits_j{\sum\limits_n {{{{e^{ - i{k_0}{\bf{\hat R}}\cdot{{\bf{r}}_j}}}{c_n}\nu_j^{(n)}}\over {{\Gamma_n}-i({\omega_k}-{\omega_{0}})}}}}}\right|^2}.\label{eq18}
\end{equation}

It is well known that the classical dipole radiation is maximal in the direction perpendicular to the dipole. Here, we have the same law for cooperative spontaneous emission which can be shown by the factor of ${\sin ^2}\theta$ in Eq.~(\ref{eq18}). What we are concerned about is not this kind of angular distribution described by ${\sin ^2}\theta $, but the directionality of spontaneous emission associated with the direction of the incident light ${{\bf{\hat k}}_0}$. So we set the detector in the x-y plane ($\theta ={\pi \mathord{\left/ {\vphantom {\pi  2}} \right.\kern-\nulldelimiterspace} 2}$) and investigate the spectra in different directions by changing the azimuthal angle $\phi$.

\begin{figure*}[b]
\includegraphics[width=4.0cm,clip]{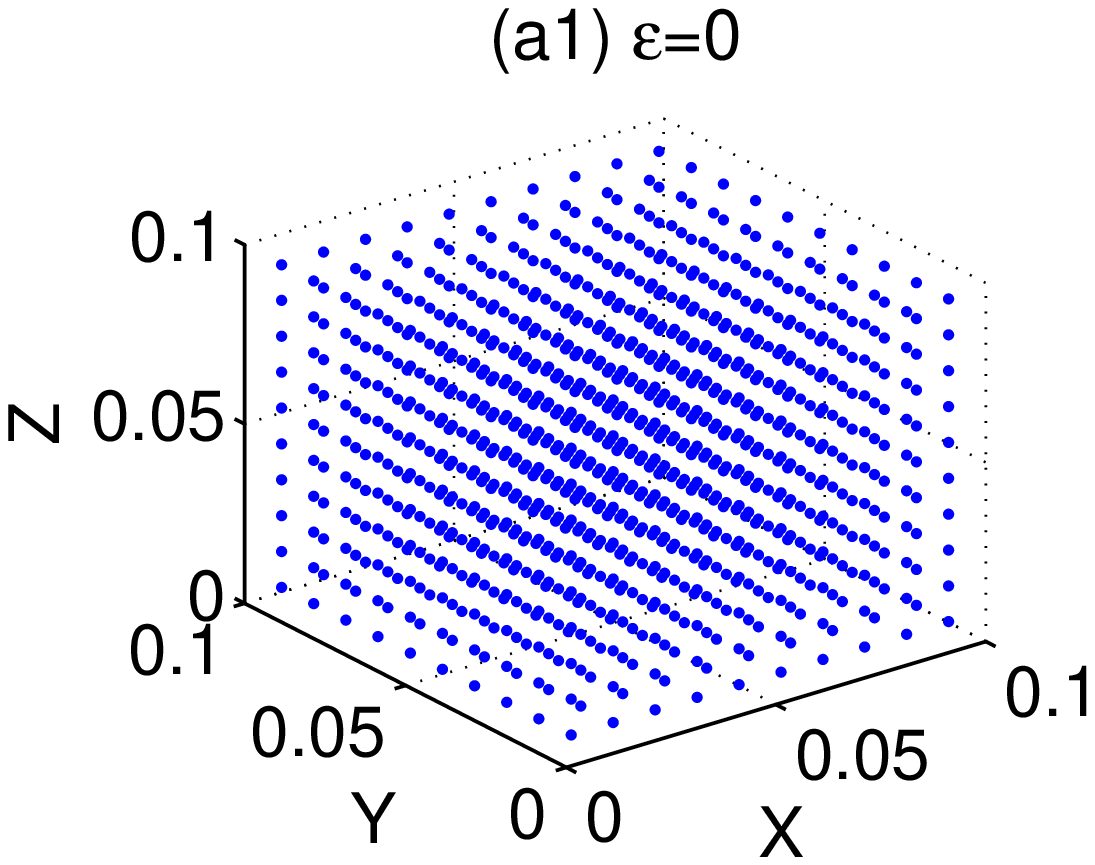}
\includegraphics[width=4.0cm,clip]{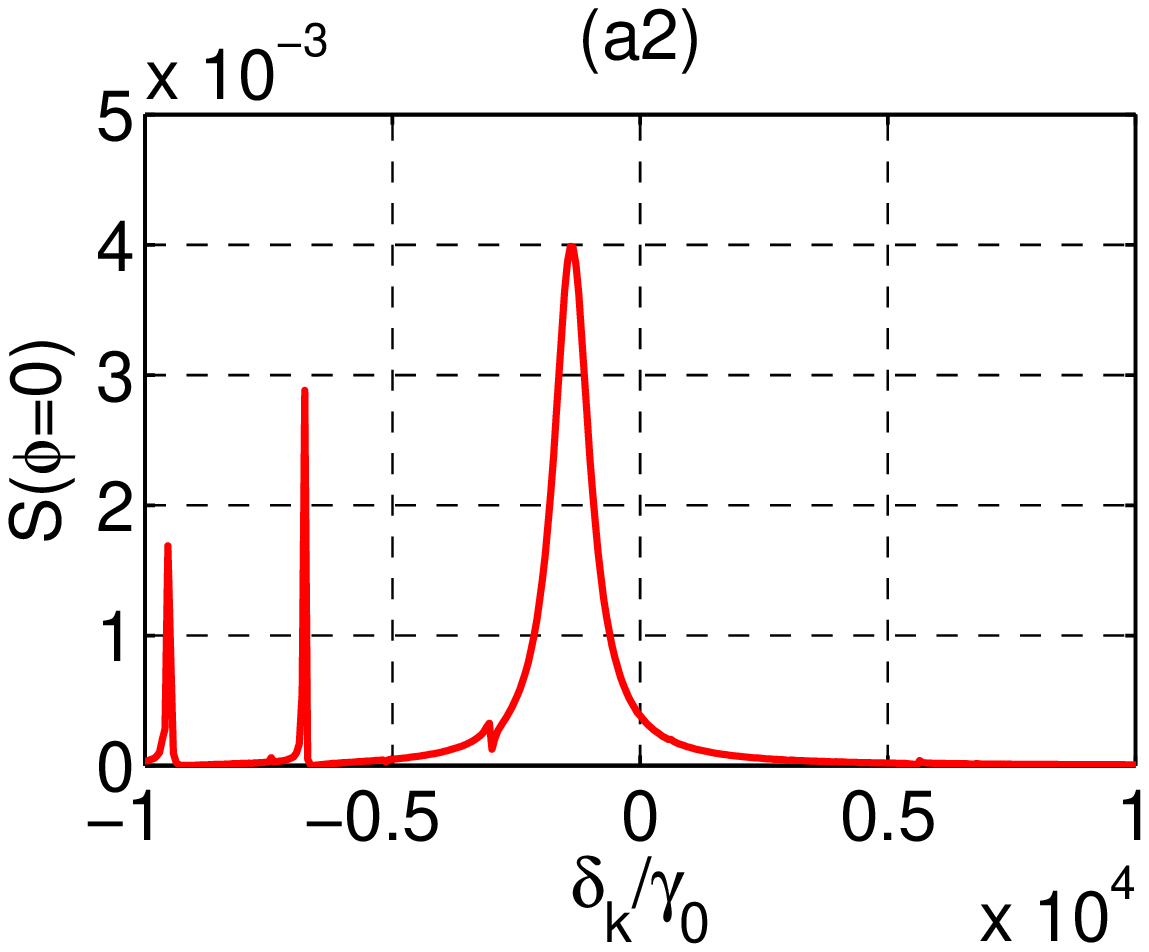}
\includegraphics[width=4.0cm,clip]{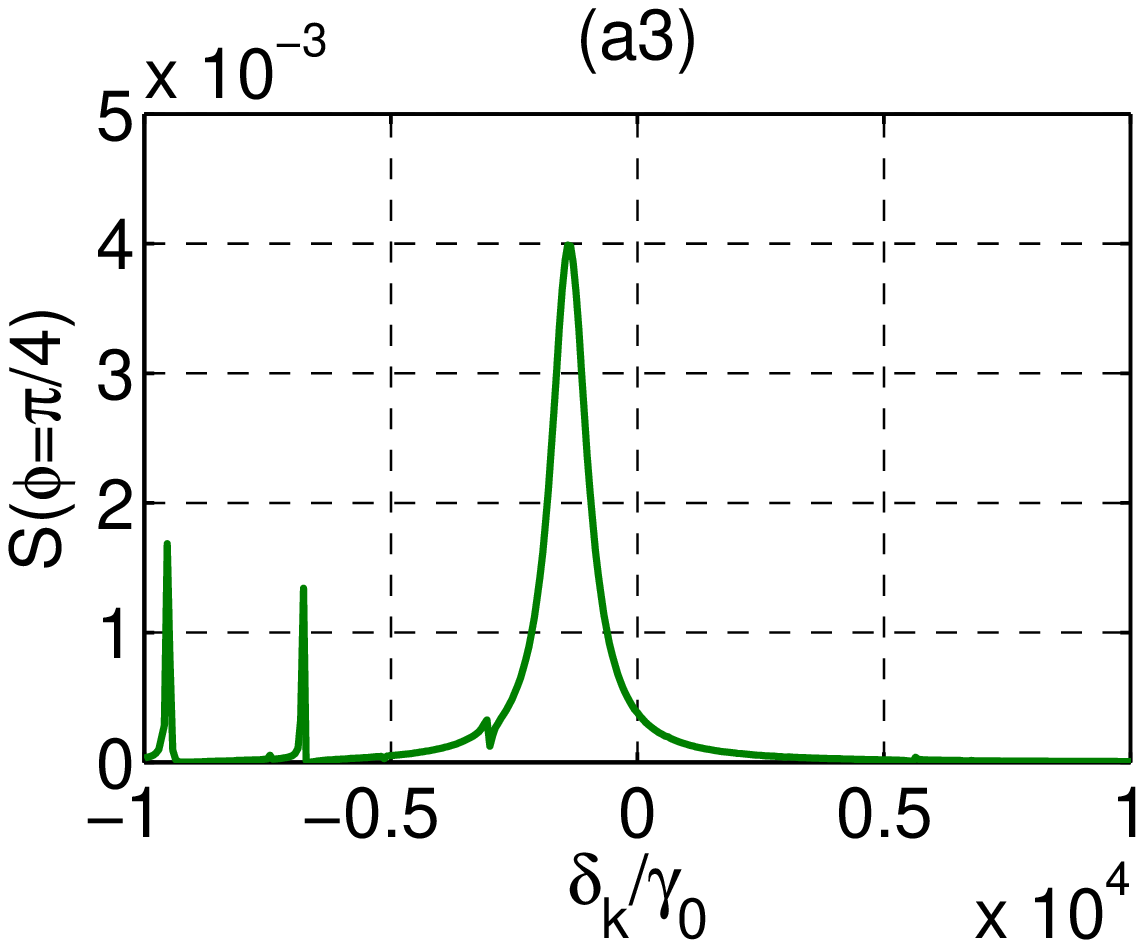}
\includegraphics[width=4.0cm,clip]{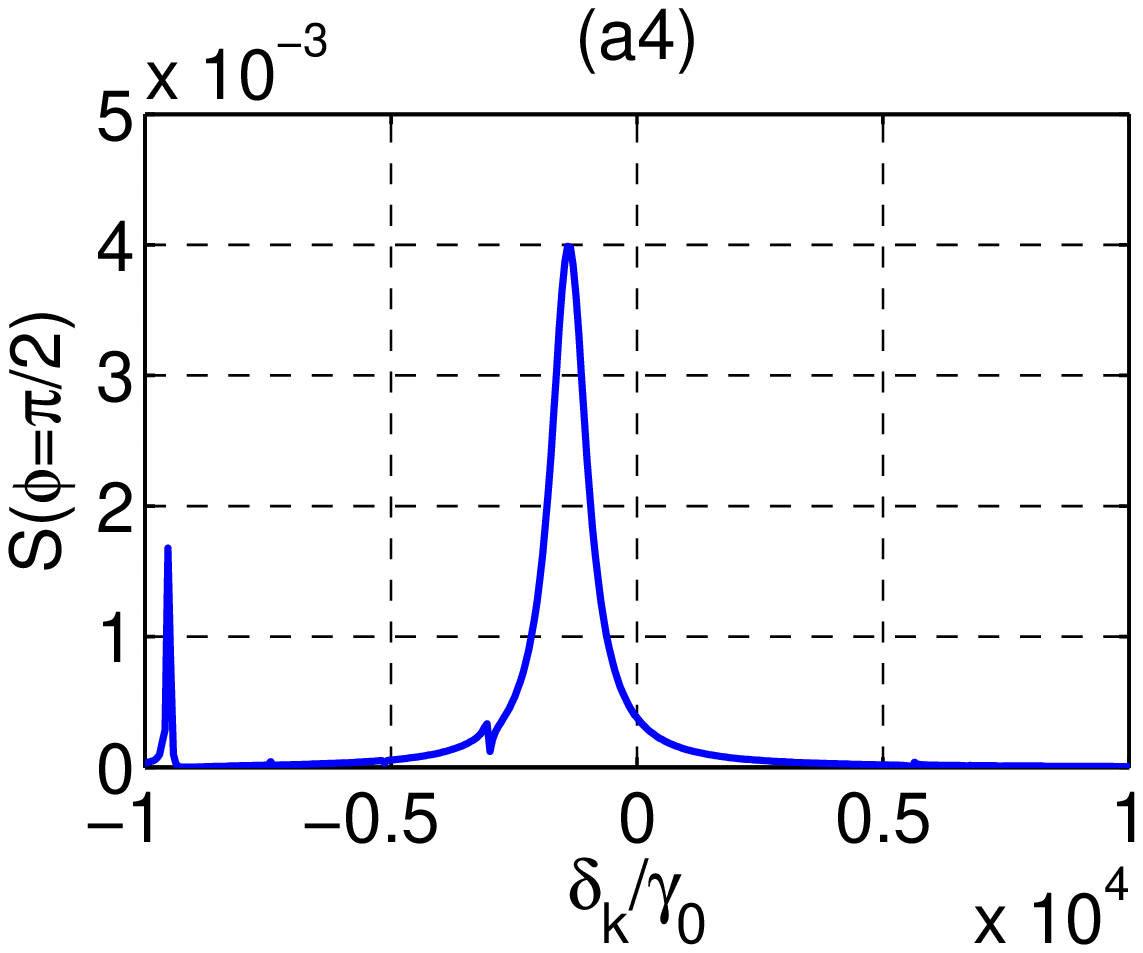}
%%%%%%%%
\includegraphics[width=4.0cm,clip]{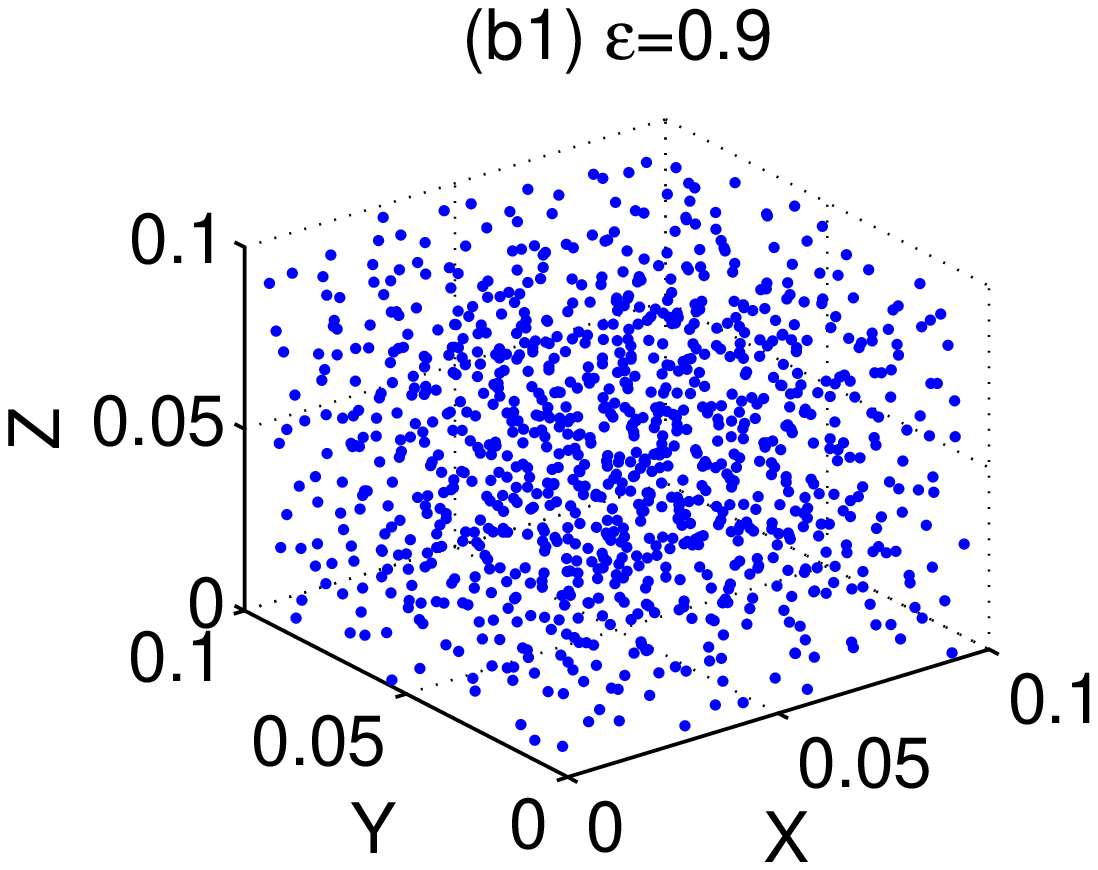}
\includegraphics[width=4.0cm,clip]{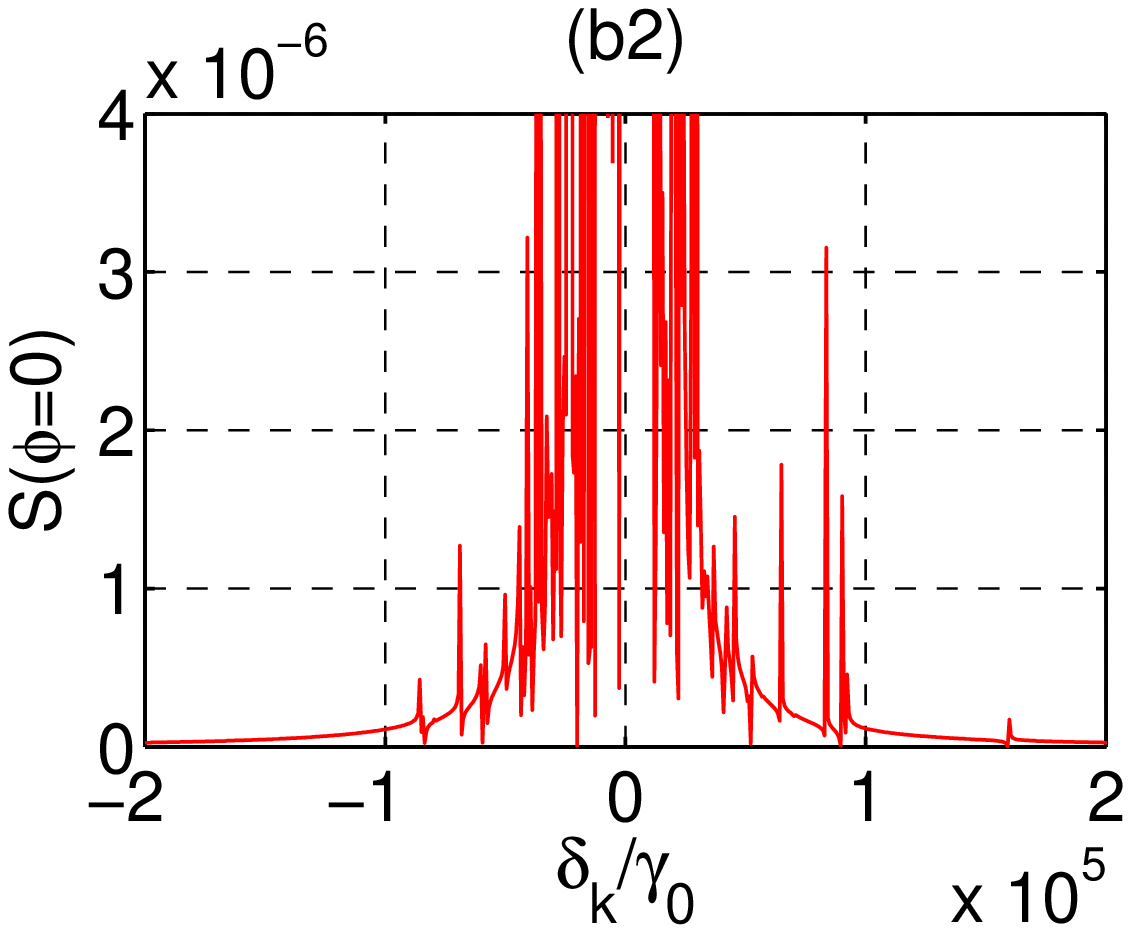}
\includegraphics[width=4.0cm,clip]{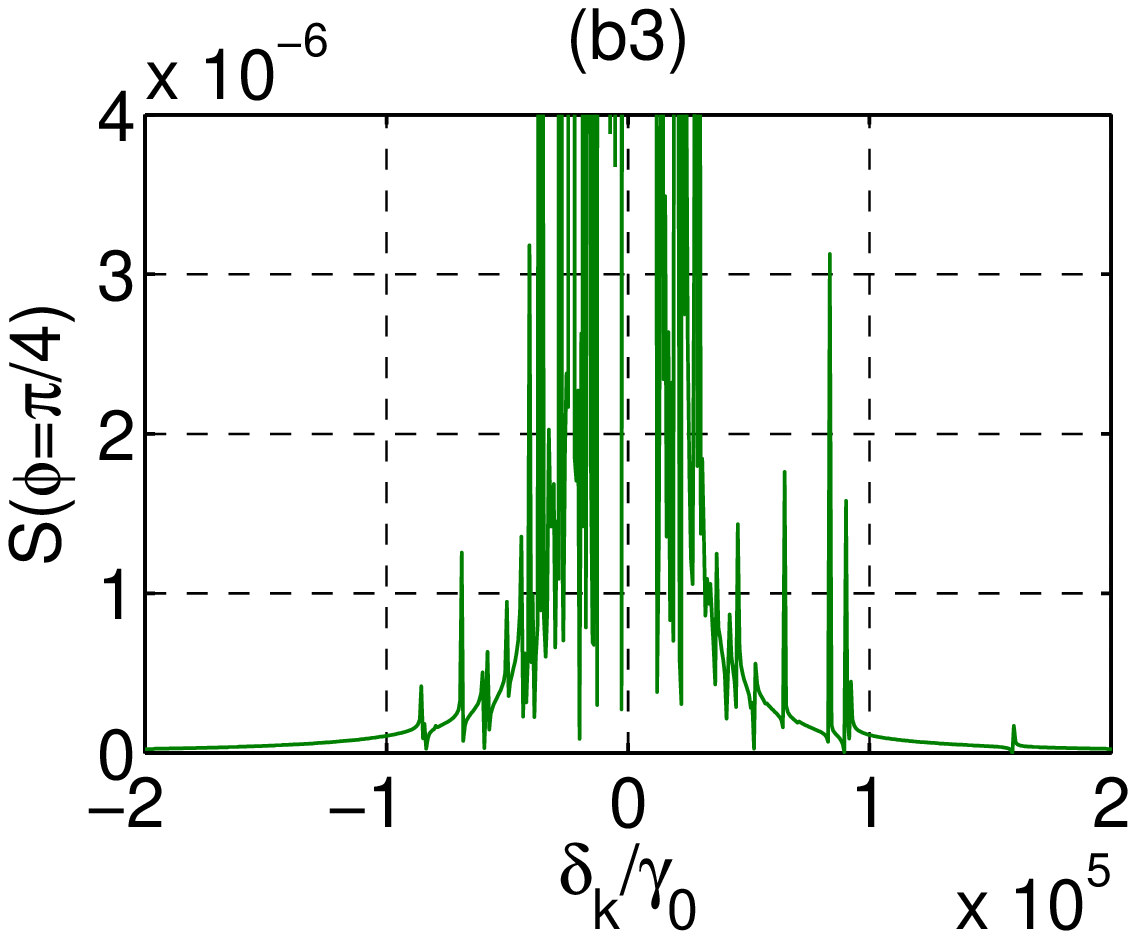}
\includegraphics[width=4.0cm,clip]{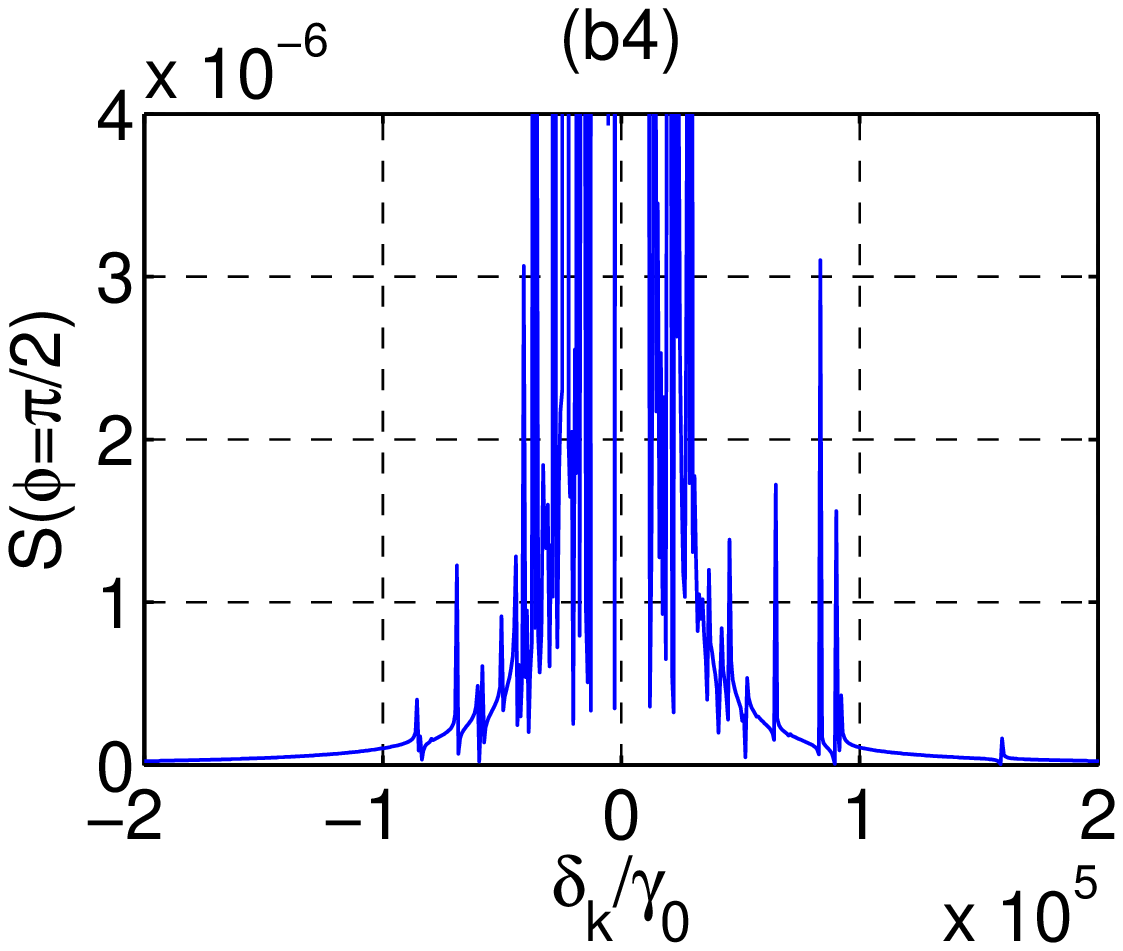}
%%%%%%
\caption{(Color online) The spectra $S(\phi)$ (in arbitrary units) in different directions with (a2) [(b2)] $\phi  = 0$, (a3) [(b3)] $\phi  = \pi /4$, and (a4) [(b4)] $\phi  = \pi /2$, corresponding to the (a1) regular [(b1) random] atomic distribution for the small sample with the side length $0.1$ (all the distances and positions are in units of $\lambda_0$). Here the incident light which prepares the initial timed Dicke state along the x-axis, the detector is in the x-y plane.}
\label{fig7}%
\end{figure*}

\begin{figure*}[htbp]
\includegraphics[width=4.0cm,clip]{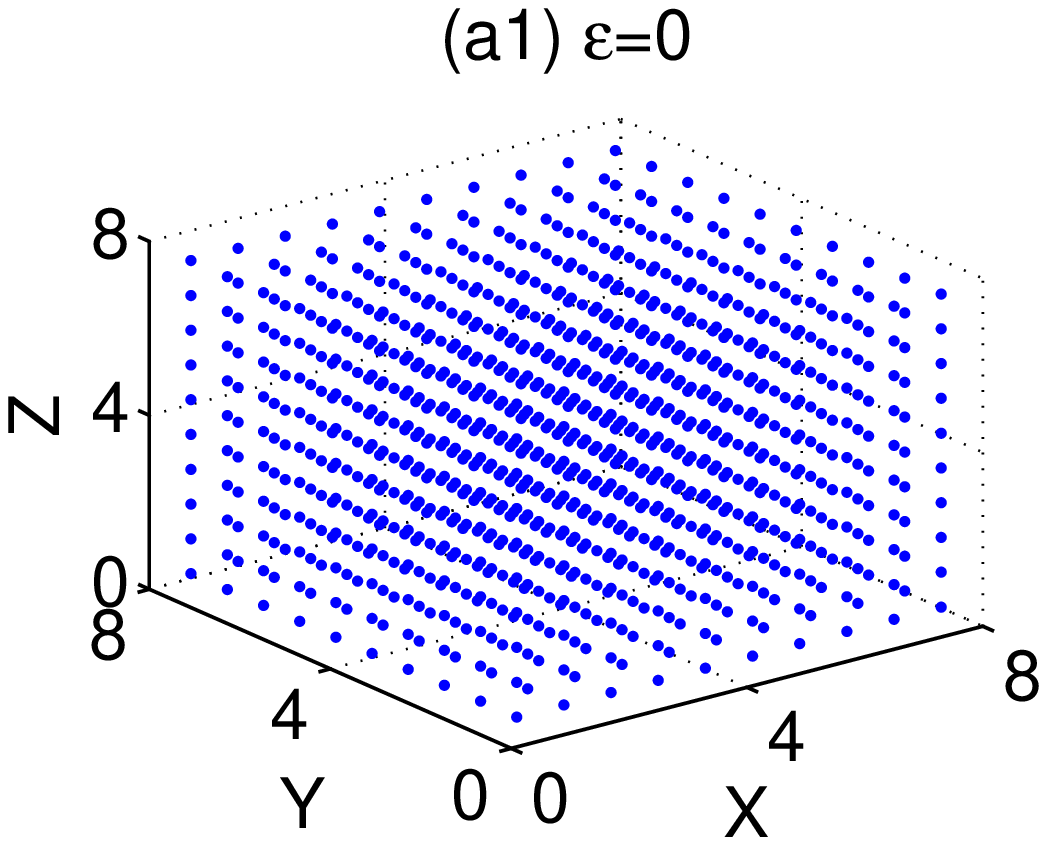}
\includegraphics[width=4.0cm,clip]{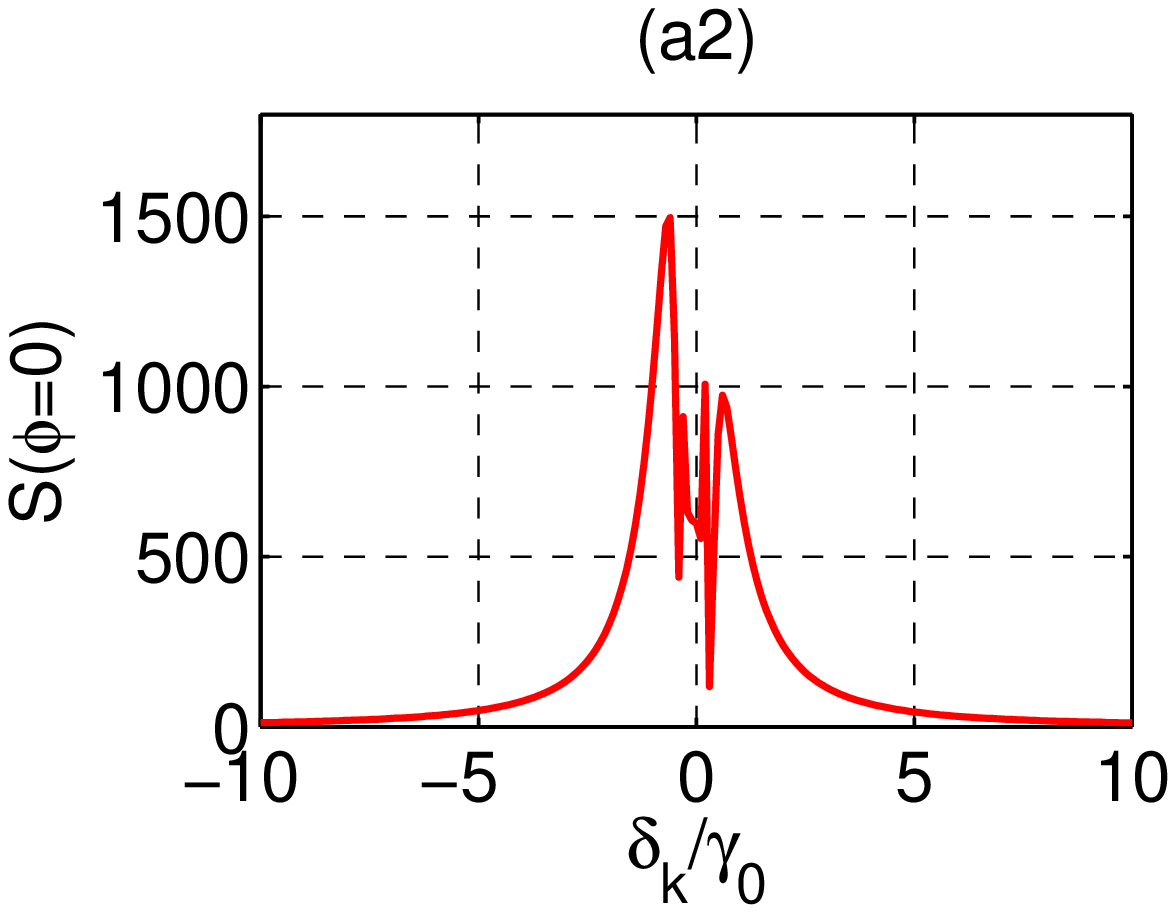}
\includegraphics[width=4.0cm,clip]{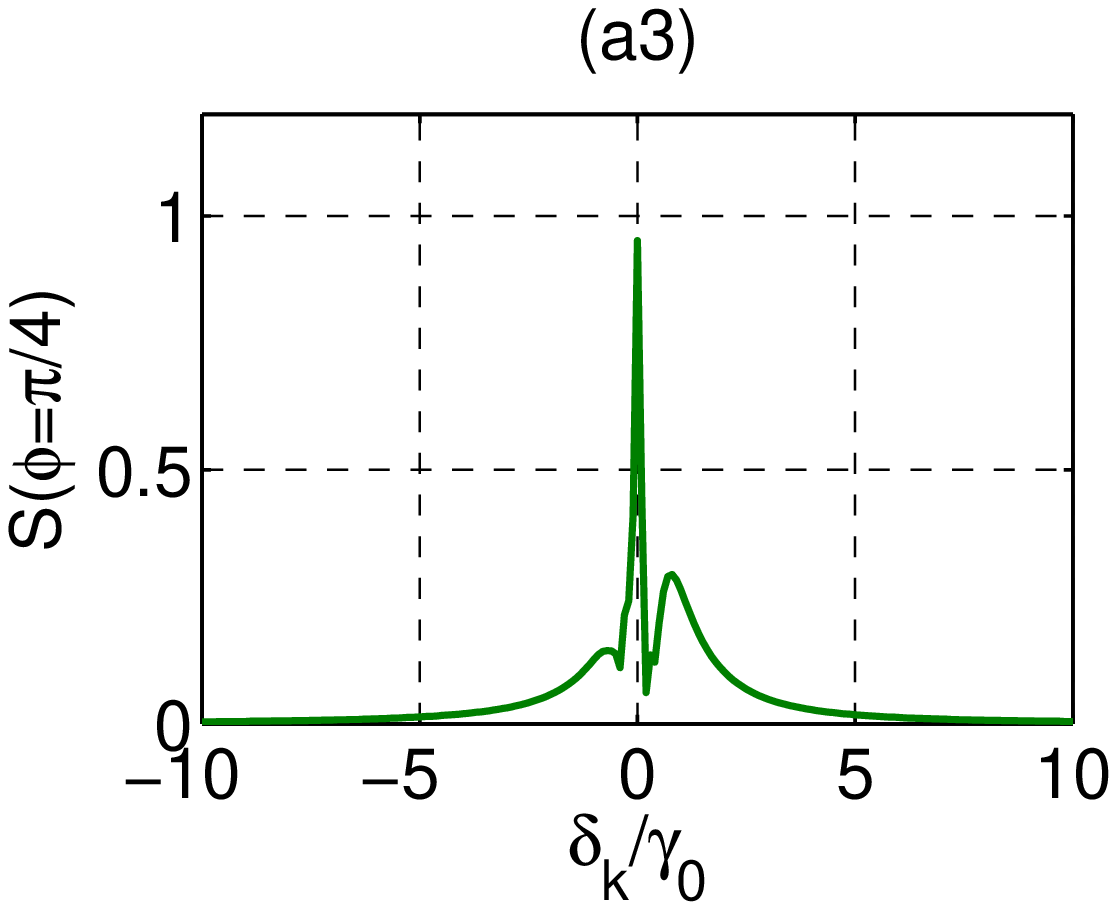}
\includegraphics[width=4.0cm,clip]{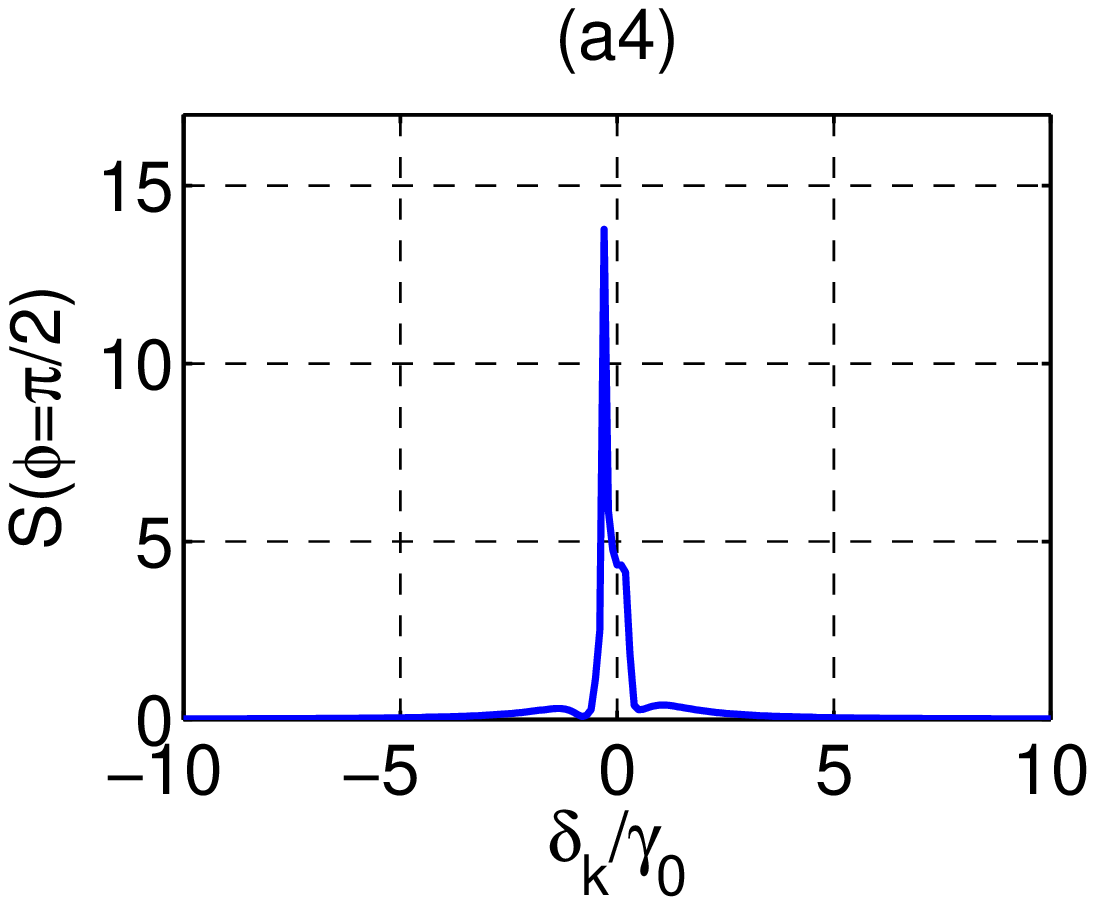}
%%%%%%%%
\includegraphics[width=4.0cm,clip]{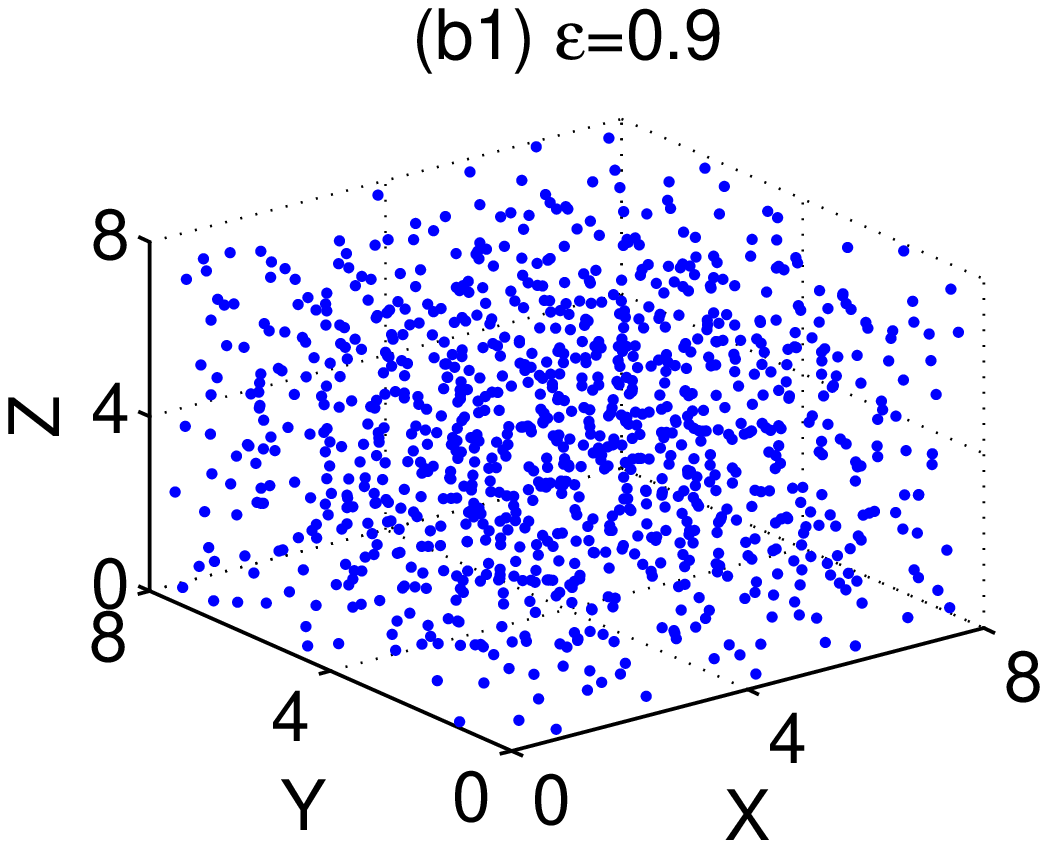}
\includegraphics[width=4.0cm,clip]{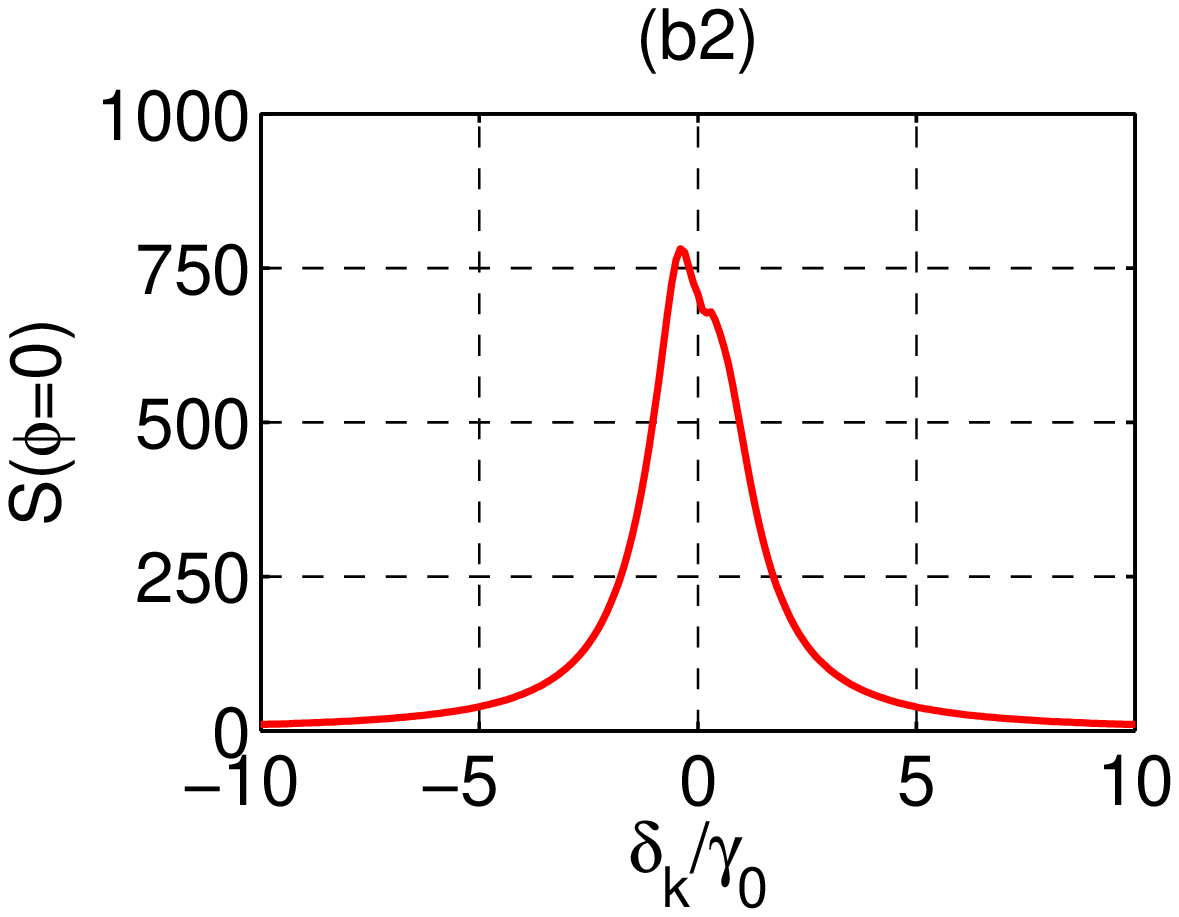}
\includegraphics[width=4.0cm,clip]{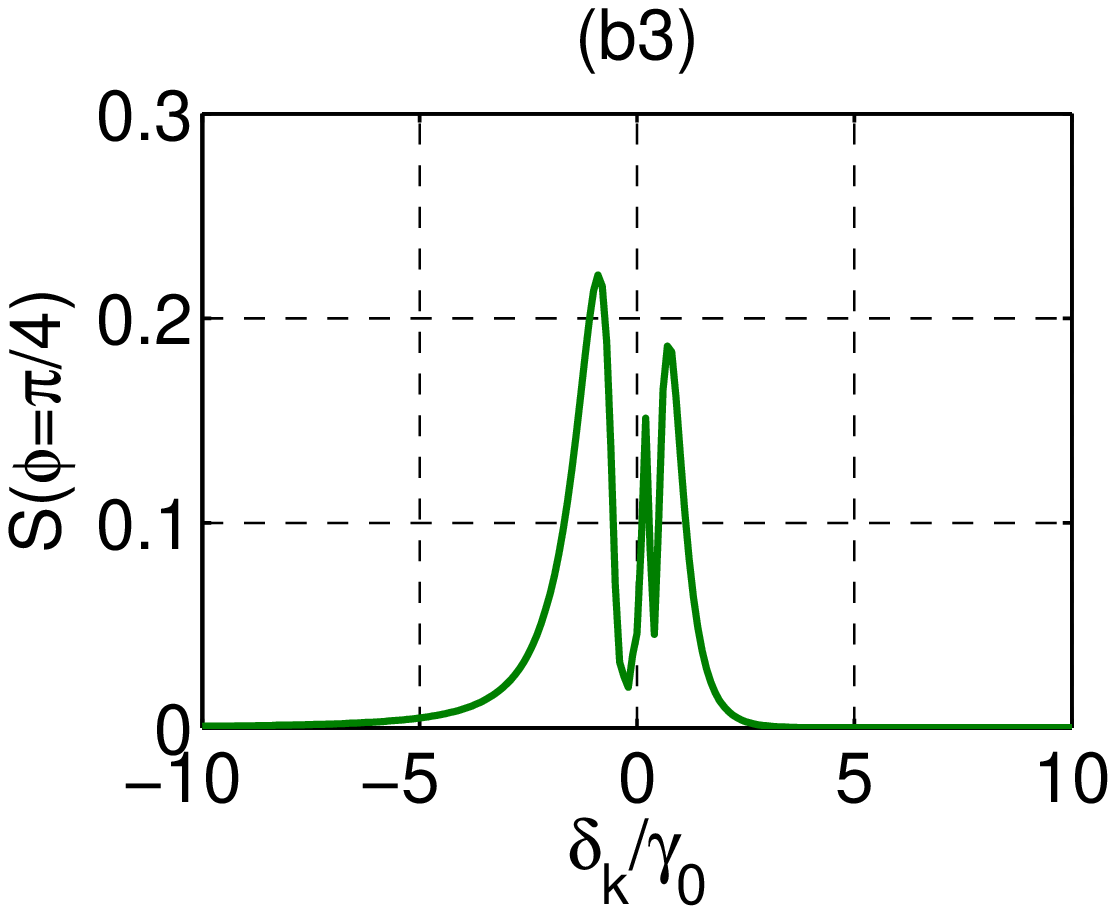}
\includegraphics[width=4.0cm,clip]{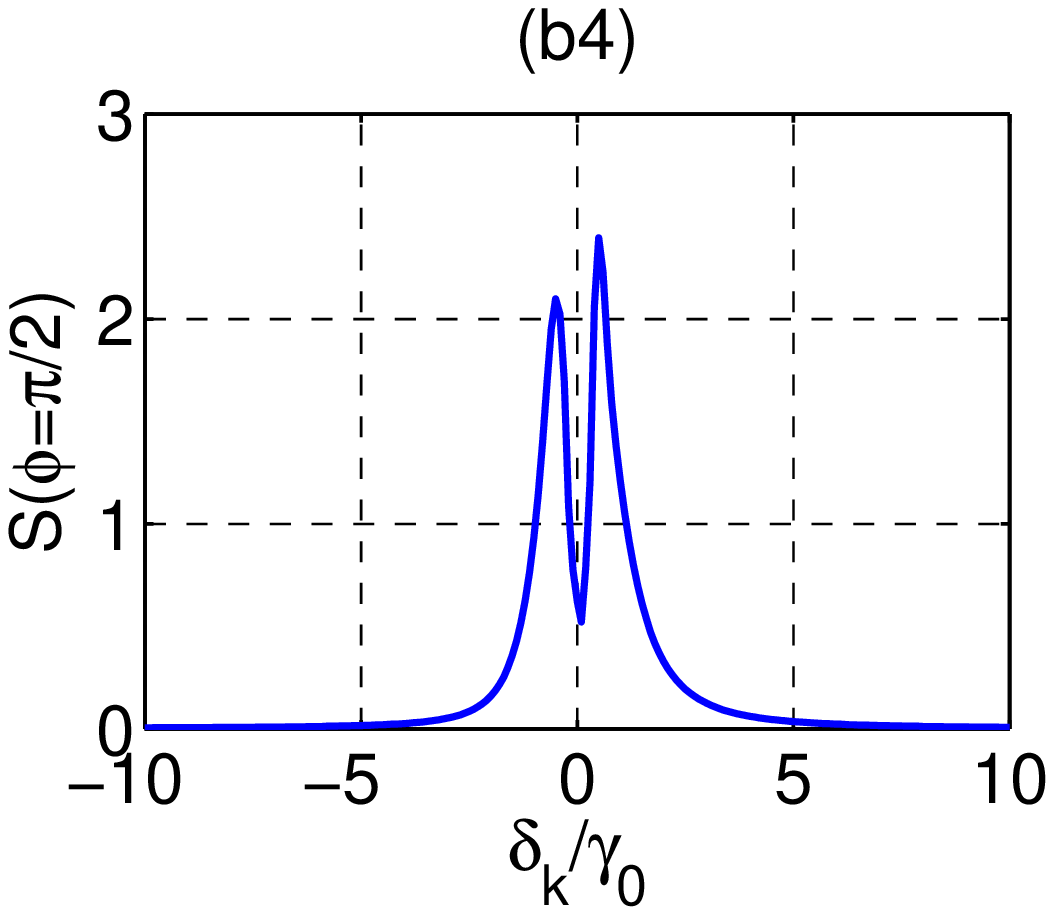}
%%%%%%%
\caption{(Color online) The spectra $S(\phi)$ (in arbitrary units) in different directions with (a2) [(b2)] $\phi  = 0$, (a3) [(b3)] $\phi  = \pi /4$, and (a4) [(b4)] $\phi  = \pi /2$, corresponding to the (a1) regular [(b1) random] atomic distribution for the large sample with the side length 8 (all the distances and positions are in units of $\lambda_0$). Here the incident light which prepares the initial timed Dicke state along the x-axis, the detector is in the x-y plane.}%
\label{fig8}
\end{figure*}

\begin{figure}[!b]
\includegraphics[width=4.0cm,clip]{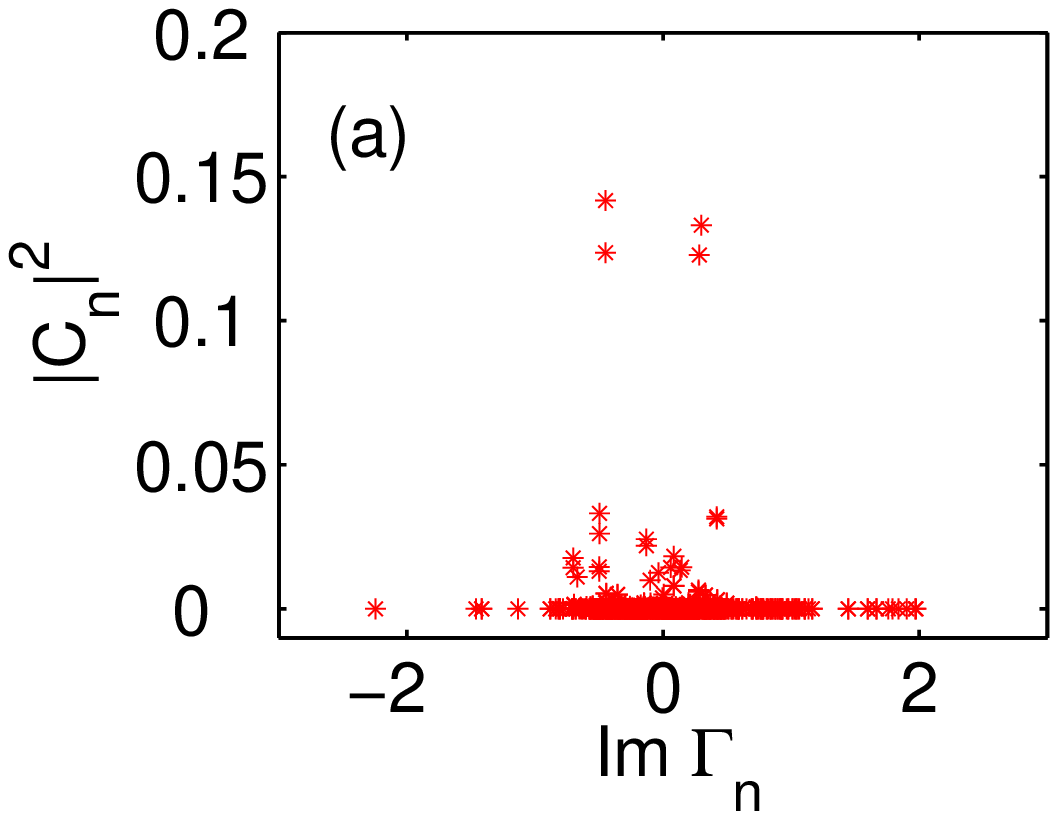}
\includegraphics[width=4.0cm,clip]{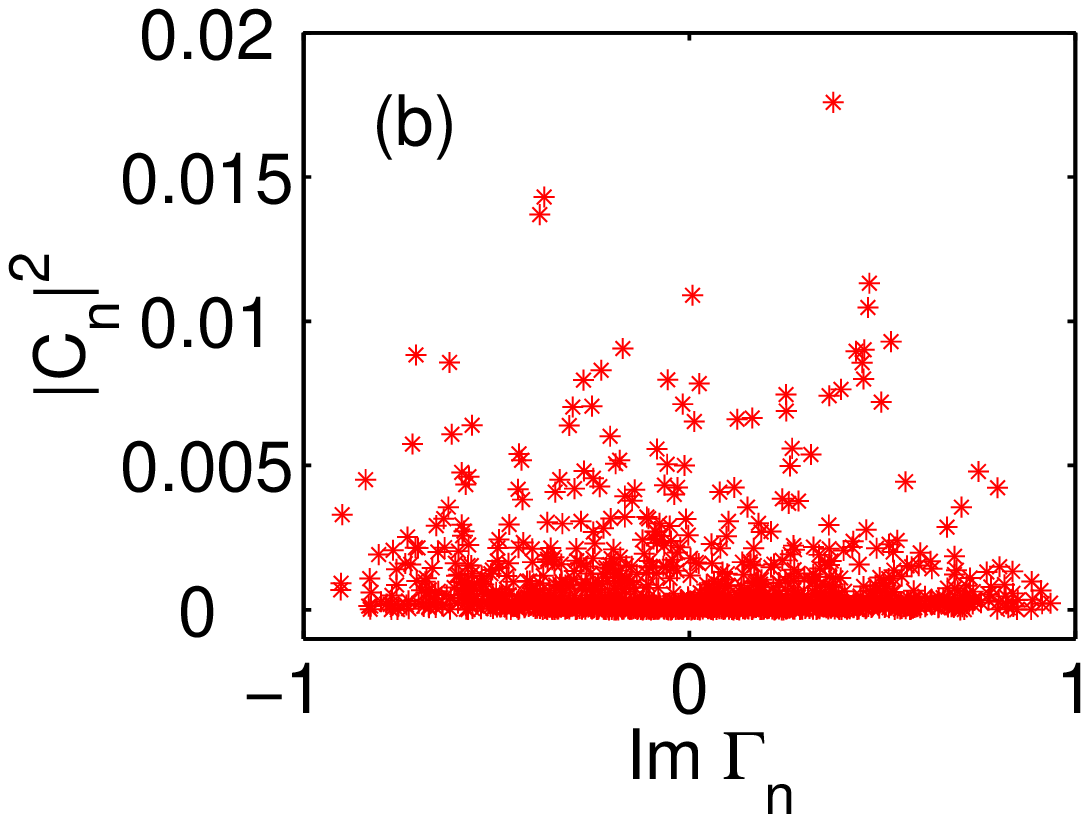}
\caption{(Color online) Magnitudes of the contribution of the different radiative eigenstates versus their corresponding Lamb shifts  ${\mathop{\rm Im}\nolimits} ({\Gamma _n})$ (in units of ${\gamma _0}$) for the (a) regular and (b) random atomic distributions as shown in Figs.~\ref{fig8}(a1) and ~\ref{fig8}(b1), respectively. The initial state is the timed Dicke state.}
\label{fig9}
\end{figure}
We plot the spectra at different angles $\phi $ for small and large samples in Figs.~\ref{fig7} and \ref{fig8}, respectively. In Fig.~\ref{fig7}, we can see that the spectra in the case of random distribution for small samples do not exhibit a distinguishable peak. Actually, it is composed of a large number of peaks corresponding to their multiple eigenstates. This can be explained by Fig.~\ref{fig2}(c2). As mentioned above, under the random atomic distribution, the Dicke state is not an approximate eigenstate but consists of many contributions associated with different eigenstates. Therefore, we obtain the spectrum shown in Figs.~\ref{fig7}(b2), (b3), and (b4). However, if the atoms are distributed regularly (uniformly), the (timed) Dicke state is an approximate eigenstate of the small-sample system [see Fig.~\ref{fig2}(a2)], so, there is a significant peak with half width about $N{\gamma_0}$ associated to the superradiant state, see Figs.~\ref{fig7}(a2), (a3) and (a4). Those additional narrow peaks come from the little components of other eigenstates. Note that there is little difference between the spectra in different directions. That means there is no directed spontaneous emission for small sample.

In Fig.~\ref{fig8}, we can clearly see the spectra in the direction of $\phi =\pi /4$ and $\phi =\pi /2$ are negligible compared with that in the direction of  $\phi  = 0$. The emission is almost along the direction of the incident light, ${{\bf{k}}_0}$. In addition, for the main direction of emission $\phi = 0$, the spectra of regular distribution has two peaks because there are two pairs of dominant eigenstates with their own similar Lamb shifts [see Fig.~\ref{fig9}(a)], while the random distribution only has one peak because the fusion of many eigenstates with quasi-continuous Lamb shifts [see Fig.~\ref{fig9}(b)].
\section{SUMMARY}
In this paper, we investigated the cooperative spontaneous emission of the system consisting of $N$ identical multilevel atoms in vacuum field by optical vector method. We focused on the influence of the details of atomic distribution, which had been ignored by most of previous papers. We find that the Dicke's original conclusion that the superradiance limit is $N{\gamma_0}$ in small samples, is only established when the atoms are distributed uniformly. Actually, in real experiments, the atoms are usually randomly distributed, and the superradiance limit $N{\gamma_0}$ will never be approached even if the dimension of the atomic sample is smaller enough than the resonance wavelength. For the large sample, the conclusion by Scully~\cite{20} that the timed Dicke state will approximately decay exponentially is correct, and the decay of symmetric Dicke state under regular distribution is slower than that in the random distribution. We also investigated the spectra, and demonstrated the directional emission for large atomic sample with the initial timed Dicke state.
\begin{acknowledgments}
This work was supported by the National Natural Science Foundation of China (under Grants No. 11174026 and No. 11174027) and the National Basic Research Program of China (Grants No. 2011CB922203 and No. 2012CB922104).
\end{acknowledgments}

\end{document}